\documentclass[apj,twocolumn,numberedappendix,twocolappendix,appendixfloats]{openjournal}
\usepackage{graphicx}
\usepackage{xspace}
\usepackage{xcolor}
\usepackage{bm}
\usepackage[ruled]{algorithm2e}
\RequirePackage[colorlinks=true,linkcolor=blue,citecolor=blue,urlcolor=blue]{hyperref}

\newcommand{\msun}{\,M$_{\odot}$\xspace}

\newcommand{\pluseq}{\mathrel{+}=}
\newcommand{\minuseq}{\mathrel{-}=}
\newcommand{\comment}[1]{\textit{{\color{black!30}#1}}}

\begin{document}

\title[]{The Synthetic Absorption Line Spectral Almanac (SALSA)\vspace{-1.5cm}}
\author{Dylan Nelson$^{1,2*}$, 
Celine Peroux$^{3,4}$, 
Philipp Richter$^{5}$, 
Matthew M. Pieri$^{4}$, 
Sebastian Lopez$^{6}$, 
Rongmon Bordoloi$^{7}$, 
Siwei Zou$^{8,6}$, 
Joseph N. Burchett$^{9}$, 
Rebecca L. Davies$^{10,11}$, 
Rahul Ramesh$^{12,1}$, 
Matthew C. Smith$^{13}$, \\
Sanchayeeta Borthakur$^{14}$, 
Christopher W. Churchill$^{9}$\vspace{0.4em}
}
\thanks{$^*$ E-mail: \href{mailto:dnelson@uni-heidelberg.de}{dnelson@uni-heidelberg.de}}

\affiliation{$^{1}$ Universit\"{a}t Heidelberg, Zentrum f\"{u}r Astronomie, ITA, Albert-Ueberle-Str. 2, 69120 Heidelberg, Germany}
\affiliation{$^{2}$ Universität Heidelberg, Interdisziplinäres Zentrum für Wissenschaftliches Rechnen, INF 205, 69120 Heidelberg, Germany}
\affiliation{$^{3}$ European Southern Observatory, Karl-Schwarzschild-Str. 2, 85748 Garching-bei-M\"{u}nchen, Germany}
\affiliation{$^{4}$ Aix Marseille Univ, CNRS, CNES, LAM, Marseille, France}
\affiliation{$^{5}$ Institut f\"{u}r Physik und Astronomie, Universit\"{a}t Potsdam, Karl-Liebknecht-Str. 24/25, 14476 Golm, Germany}
\affiliation{$^{6}$ Departamento de Astronomía, Universidad de Chile, Casilla 36-D, Santiago, Chile}
\affiliation{$^{7}$ Department of Physics, North Carolina State University, Raleigh, 27695, North Carolina, USA}
\affiliation{$^{8}$ Chinese Academy of Sciences South America Center for Astronomy, National Astronomical Observatories, Beijing 100101, China}
\affiliation{$^{9}$ Department of Astronomy, New Mexico State University, Las Cruces, NM 88003, USA}
\affiliation{$^{10}$ Centre for Astrophysics and Supercomputing, Swinburne University of Technology, Hawthorn, Victoria 3122, Australia}
\affiliation{$^{11}$ ARC Centre of Excellence for All Sky Astrophysics in 3 Dimensions (ASTRO 3D), Australia}
\affiliation{$^{12}$ Kavli IPMU (WPI), UTIAS, The University of Tokyo, Kashiwa, Chiba 277-8583, Japan}
\affiliation{$^{13}$ Max-Planck-Institut f\"{u}r Astrophysik, Karl-Schwarzschild-Str. 1, D-85748, Garching, Germany}
\affiliation{$^{14}$ School of Earth \& Space Exploration, Arizona State University, 781 Terrace Mall, Tempe, AZ 85287, USA}

\begin{abstract}
We create the first large-scale mock spectroscopic survey of gas absorption sightlines traversing the interstellar medium (ISM), circumgalactic medium (CGM), and intergalactic medium (IGM) surrounding galaxies of virtual Universes. That is, we create mock, or synthetic, absorption spectra by drawing lines-of-sight through cosmological hydrodynamical simulations, using a new mesh-free Voronoi ray-tracing algorithm. The result is the Synthetic Absorption Line Spectral Almanac (SALSA), which is publicly released on a feature-rich online science platform.\thanks{\url{www.tng-project.org/spectra}} It spans a diverse range of ions, transitions, instruments, observational characteristics, modeling assumptions, redshifts, and numerical simulations. These include, but are not limited to: (ions) HI, OI, CI, MgI, MgII, FeII, SiII, CaII, ZnII, SiIII, SiIV, NV, CII, CIV, OVI; (instruments) SDSS-BOSS, KECK-HIRES, UVES, COS, DESI, 4MOST, WEAVE, XSHOOTER; (model choices) with/without dust depletion, noise, quasar continua, foregrounds; (redshift) from $z=0$ to $z \simeq 6$; (ancillary data) integrated equivalent widths, column densities, distances and properties of nearby galaxies; (simulations) IllustrisTNG including TNG50, TNG-Cluster, EAGLE, and SIMBA. We also provide line-of-sight integrals, including Faraday rotation measure (RM) probing magnetic fields, and dispersion measure (DM) as traced by fast radio bursts (FRBs). This scope is not fixed, and will grow and evolve with community interest and requests over time. The resulting dataset is generic and broadly applicable, enabling diverse science goals such as: (i) studies of the underlying physical gas structures giving rise to particular absorption signatures, (ii) galaxy-absorber and halo-absorber correlations, (iii) virtual surveys and survey strategy optimization, (iv) stacking experiments and the identification of weak absorption features, (v) assessment of data reduction methods and completeness calculations, (vi) inference of physical properties from observables, and (vii) apples-to-apples comparisons between simulations and data.
\end{abstract}






\section{Introduction}

The circumgalactic medium (CGM) and intergalactic medium (IGM) are key components of the cosmic ecosystem, and play pivotal roles in the evolution of galaxies. Together, the CGM and IGM are by far the dominant baryon reservoirs in the Universe. They contain all the diffuse gas residing outside of galaxies, and mediate the interplay between accretion i.e. inflows and feedback-driven outflows. As a result, they provide the fuel for star formation in galaxies, regulate feedback, and create many unique observational signatures of galaxy formation in action. 

Despite their importance, the CGM and IGM remain poorly understood due to (i) the observational challenges of detecting such low density gas and (ii) the difficulty of interpreting observed features. One of the most important observational tools is absorption line spectroscopy, particularly at rest-frame optical and ultraviolet (UV) wavelengths, where gas at densities and temperatures characteristic of the CGM and IGM has numerous atomic transitions, in both hydrogen and metals \citep{tumlinson17}. Absorption spectra characterize the light blocked by any intervening media towards a luminous background source, be it a quasar, galaxy, star, or GRB.

Due to their abundance and relatively simple intrinsic continua, bright quasar sightlines are a workhorse of CGM/IGM studies \citep{rauch98,peroux20a}, particularly due to their large numbers in all-sky surveys such as SDSS and DESI \citep{zhu13}. In particular, the Lyman-$\alpha$ forest, a series of absorption lines caused by neutral hydrogen in the IGM, has been instrumental in mapping the distribution of matter in the Universe \citep{sargent80}, inferring cosmological parameters \citep{mcdonald08,busca13,fernandez24}, understanding the physics of the IGM \citep{bolton17,puchwein23}, and constraining the nature of dark matter itself \citep{viel05,seljak06}. Similarly, hydrogen and metal absorption lines, such as those from C IV, O VI, and Mg II, offer insights into the enrichment history and galactic feedback processes that shape the CGM \citep{chen98,peeples14,lan18}.

We can study absorption in several contexts. First, blind quasar i.e. absorption-selected surveys look exclusively at the occurrence of absorption as a function of redshift, leading to summary statistics such as the column density distribution function and comoving line density \citep{zou21,davies23,yu25}, as well as pixel optical-depth methods \citep{aguirre02,burchett20}. Follow-up spectroscopic campaigns or large-field IFU observations can subsequently identify foreground galaxies at close projected separations \citep{nielsen13,schroetter16,chen20,dutta20,weng23,beckett24}.

Alternatively, a galaxy-centric approach starts with a well-defined population of foreground galaxies without any prior knowledge of their CGM \citep{steidel10,tumlinson11,bordoloi14c}. These galaxy-selected surveys allow CGM constraints as a function of galaxy properties, from stellar mass to orientation \citep{bordoloi11,kacprzak15,werk16}. For a sufficiently deep galaxy follow-up campaign on absorption-selected quasar sightlines, these two approaches converge \citep{higginson25}.

The CGM of our own Milky Way also offers unique information, from the resolved structure of high-velocity clouds \citep{muller63} to variation and coherence between closely spaced sightlines \citep{howk02,richter17,bish19}. In addition, large optical surveys such as SDSS and DESI obtain both galaxy and quasar samples, now numbering into the tens of millions. This enables statistical studies of absorption with respect to (low-redshift) foreground galaxy properties \citep{zhu14,anand21,chen25} as well as cross-correlations with multi-wavelength surveys \citep{chang24,zou24}. Stacking experiments with large datasets can also reveal intrinsically faint absorption and probe otherwise inaccessible phases \citep[e.g.][]{york83,richter25}.

Absorption spectra provide a unique probe of the gas in and around galaxies, revealing information about its ionization state, metallicity, temperature, and kinematics \citep{fg23}. However, interpreting such data is challenging.

First, this is because of the underlying physical complexity: gaseous halos are intrinsically multi-scale, multi-phase, and multi-physics. The CGM contains structure across an enormous range of scales \citep{rauch99,rigby02,pieri14,crighton15}, from halo spanning bulk flows to sub-parsec scale cool clouds and interface layers \citep{fielding20,nelson20,ramesh24a,augustin25}. As a result, halo gas co-exists across a broad range of temperatures and experiences rapid mass exchange between phases \citep{mccourt12,beckmann19,ramesh25b}. Finally, many physical processes modulate the CGM, with coupled and non-linear interactions between cooling and chemistry including dust \citep{richings16,aoyama18}, turbulence \citep{ruszkowski10,vazza11b,fournier25}, non-thermal components including magnetic fields and cosmic rays \citep{ji20,pakmor20,ruszkowski23}, galactic outflows \citep{suresh15,mitchell18,pillepich21,rey24}, all in the presence of time and space-varying radiation fields \citep{kannan16,oppenheimer18a,obreja24}.

Second, the interpretation of absorption line spectroscopy is also made challenging by the inherently limited view offered by the data itself. For example, while a single sightline has line-of-sight kinematic resolution given by its spectral resolution, it has no spatial resolution on the plane of the sky. Exceptions include spatially extended background sources such as resolved galaxies \citep{rubin18,peroux18}, multiply lensed quasars \citep{rauch99}, gravitationally lensed arcs \citep{lopez18,bordoloi22}, and dense multi-sightline sampling of nearby halos \citep[e.g. the AMIGA survey of M31;][]{lehner20}. In addition, the ability to resolve in velocity space is complicated by narrow components with small velocity offsets that can blend and so appear as a single absorption line at low spectral resolution \citep{tripp96,linsky25}. That is, small-scale line of sight structure is lost due to line broadening, line blending, and redshift space distributions. As a result, modeling absorption signatures arising from the CGM and IGM is a non-trivial process with many simplifying assumptions, caveats, and possible pitfalls \citep{marra21,marra24,sameer24,taira25}.

One of the most powerful theoretical tools to tackle these challenges are large-volume cosmological magnetohydrodynamical simulations of galaxy formation and evolution \citep{naab17}. These numerical models are now a cornerstone of theoretical astrophysics. Projects such as Illustris, EAGLE, Horizon-AGN, IllustrisTNG, and Simba provide `virtual universes' to study the formation and evolution of galaxies and their surrounding environments \citep{nelson19a}. Hydrodynamical simulations in particular self-consistently combine self-gravity, gas dynamics, and small-scale processes such as radiative cooling, star formation, and feedback from stars and supermassive black holes \citep{vog20}. Although caveats exist, overall these simulations generate reasonably realistic representations of gas in the CGM and IGM, as demonstrated through numerous quantitative comparisons \citep[e.g.][]{ford13,bird14,fg16,rahmati16,nelson18b,oppenheimer18c,appleby21,amodeo21,zhang25}. They also produce realistic galaxy populations, enabling the study of connections and correlations between gas absorption and nearby galaxies and halos.

However, making the link between numerical simulations and absorption spectroscopy is non-trivial, and is a much needed and often missing step. Synthetic absorption spectra derived from cosmological simulations are perhaps the most robust and effective way to bridge the gap between theory and observations \citep{peroux24}. This simultaneously enables: (i) robust, apples-to-apples comparisons and validations of the theoretical models; (ii) interpretation of observational data and inferences of the underlying physical gas properties that give rise to absorption signatures. The latter are particularly important, as simulations have access to the full three (six) dimensional distribution of gas (and its kinematics), as well as quantities such as total hydrogen number density and total metallicity that are inaccessible with single metal-line tracers. They also provide time evolution, evolutionary history, statistical samples, and the ability to probe and test the role of different physical processes.

These details can be obscured in observations due to projection effects \citep{ho21}, noise, finite spectral resolution, modeling degeneracies, and so on. Because of the loss of information, the inference of physical gas properties from observed spectra is an inherently degenerate process \citep{sameer21,marra24} -- the inverse problem is not one-to-one \citep{hafen24}. In the other direction, however, `forward modeling' of a given simulated gas distribution into its observable absorption spectrum is well-posed and has no ambiguity, other than the inherent assumptions and choices involved in theoretical models.

Several methods exist for generating synthetic absorption spectra. The \textsc{pygad} package creates spectra from SPH simulations, avoiding any spatial binning or smoothing in favor of particle by particle modeling \citep{rottgers20}. \textsc{fsfe} operates with a similar, particle-by-particle or cell-by-cell approach \citep{bird17}. In contrast, \textsc{specexbin} bins SPH particles along sightlines before photoionization modeling and spectral generation \citep{oppenheimer06,dave10,ford13}. Focused more on Eulerian simulation outputs, the \textsc{trident} code is also publicly available and uses \textsc{yt} to load and represent gas distributions, making it by far the most widely used \citep{hummels17}. The \textsc{specwizard} code was originally developed for the Ly-$\alpha$ forest \citep{theuns98} and subsequently extended for metal-line absorption \citep{teppergarcia11,wijers19}. Particularly in the Ly-$\alpha$ forest community, many authors have developed similar techniques without necessarily giving them fun names \citep[e.g.][]{cen94,hernquist96,haehnelt96,shen13,keating14,egan14,hennawi21}. For metal ionization, essentially all approaches use the \textsc{Cloudy} photoionization code \citep{ferland17,gunasekera25} in post-processing, and thus incur assumptions related to equilibrium \citep[avoided in on-the-fly chemistry models such as][]{richings14,katz22}. Some adopt approximate self-shielding models to account for attenuation of the background radiation field in denser environments \citep{rahmati13}.

Simulation-based mock spectra have been applied to several problems. These include assessing how, and when, the physical properties of CGM gas can be successfully inferred from absorption lines \citep{liang16}; understanding the degree of line-of-sight blending from kinematically coincident but spatially distinct gas structures \citep{churchill15}; comparing velocity structure within absorbers to observations \citep{peeples19}; identifying the origin of gas giving rise to particular absorption features \citep{emerick15}; contrasting statistics derived from realistic mocks versus directly extracted simulation values \citep{egan14}; understanding how close sightline pairs can constrain small-scale cloud structure (\textcolor{blue}{Guo et al. in prep}); interpreting the kinematics of absorbing gas \citep{kacprzak19}; and quantitatively comparing models with data \citep{vandervliet17,marra21}.

Providing new analysis techniques, machine learning approaches trained on mock spectra can be used to identify absorbers \citep{szakacs23} and create mappings between absorption observables and physical gas properties \citep{appleby24}, treating hydrodynamical simulations as ground truth training datasets \citep[see also][]{stemock24,jalan24,pistis25}. Beyond low-redshift, mock absorption spectra are also crucial at other scales and epochs. At high-redshift, they can quantify key Ly$\alpha$ forest statistics \citep{chabanier23}, infer the progress of cosmic reionization from Ly$\alpha$ forest transmission spikes \citep{garaldi19}, constrain the sizes of cosmic HII bubbles surrounding galaxies \citep{nikolic25}, and identify the origin of double-peaked Ly$\alpha$ signatures in high-redshift galaxies \citep{hu16}, as seen in recent JWST data.

\begin{figure*}
   \centering
   \includegraphics[width=1.0\textwidth]{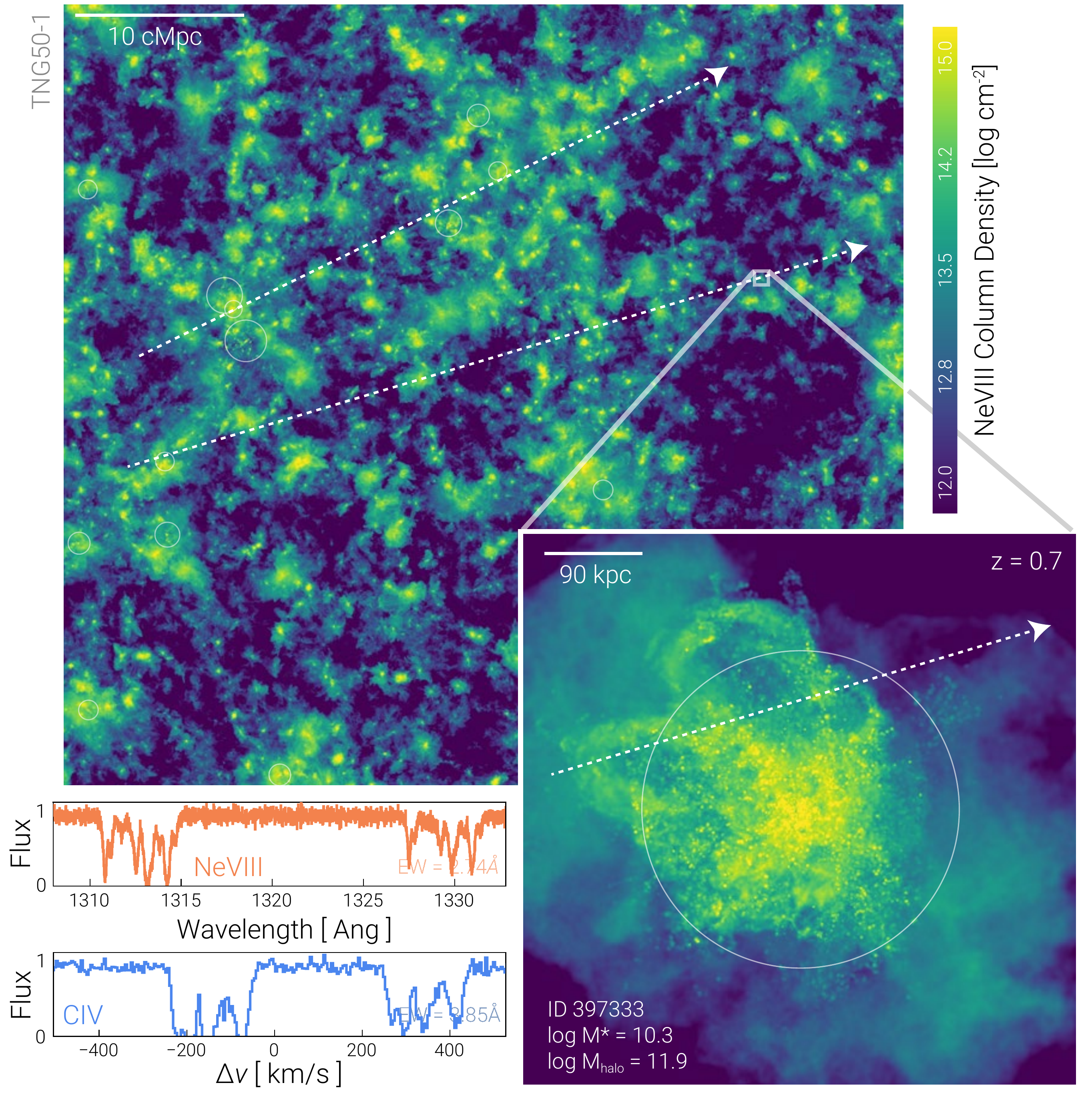}
   \caption{Schematic overview of the creation of synthetic absorption spectra from a cosmological hydrodynamical simulation. The main view shows projected NeVIII column density at $z=0.7$ across the TNG50 volume. Two sightlines traverse this simulated volume (dashed white line); one intersects the projected virial radius (white circle) of a galaxy with $M_\star = 10^{10.3}$\msun and $M_{\rm halo} = 10^{11.9}$\msun, as shown in the lower-right inset. The circumgalactic medium of this halo shows a complex morphology, with bubble-like features driven by galactic feedback. The resulting density and kinematic structure is imprinted in observable spectroscopic signatures such as the NeVIII 780, 770\AA\xspace doublet (mock COS-G130M; orange, inspired by the CASBaH survey of \protect\citealt{burchett19}). The second sightline intersects different gas that produces corresponding CIV 1548, 1550\AA\xspace doublet absorption (mock HIRES-B14, at $z=2$; blue). }
   \label{fig:vis_overview}
\end{figure*}

Finally, beyond the optical/UV regime, absorption of diffuse gas also plays a key role in the detection and characterization of the CGM and warm-hot intergalactic medium (WHIM) via X-ray absorption lines \citep{nicastro18,das21,mathur22}.  The dispersion measure of fast radio bursts (FRBs) offers a similar, line-of-sight integral probe of low-density ionized gas in the IGM and CGM \citep[e.g.][]{prochaska19b}.

Myriad uses of mock absorption spectra exist. However, creating synthetic sightlines and spectra from cosmological galaxy formation simulations can be challenging, especially for high-resolution simulations and/or large numbers of spectra. The majority of the studies above create and use tailored datasets targeting specific questions. Few libraries of synthetic spectra have been released or are publicly available for use. Notable exceptions include \citet{garel24} who present a library of $\sim 50,000$ synthetic Ly$\alpha$ and metal line emission outflow spectra,\footnote{\url{rascas.univ-lyon1.fr/app/idealised_models_grid/}} although these are based on idealized (1D) shell models and not galaxy formation simulations. \citet{byrohl20} release a library of $\sim 500,000$ Ly$\alpha$ transmission curves through the IGM derived from TNG100.\footnote{\url{zenodo.org/records/3832098}} There are several catalogs of Ly$\alpha$ forest spectra available, mostly for cosmology purposes. From hydrodynamical simulations, \citet{tilman23} create and release $5,000$ Ly$\alpha$ forest spectra for each simulation of the CAMELS suite. However, as far as we are aware, no general purpose library of synthetic absorption sightlines based on cosmological simulations exists, or is available.

In this work, we describe the creation of a new synthetic absorption spectra generation tool, tailored to cosmological hydrodynamical galaxy formation simulations. We then apply this tool to several existing and public large-volume simulations in order to create, and publicly release (\url{www.tng-project.org/spectra}) a large library of synthetic absorption spectra. Figure \ref{fig:vis_overview} gives a visual schematic of the project and the resulting data products.

The rest of this paper is organized as follows: Section \ref{sec_methods} presents our methods: ionization modeling (Section \ref{sec_ions}), abundances and dust (Section \ref{sec_abunddust}), creating sightlines and our new mesh-free Voronoi ray-tracing algorithm (Section \ref{sec_sightlines}), generating mock spectra (Section \ref{sec_mockspec}), and adding instrumental effects and realism (Section \ref{sec_inst}). In Section \ref{sec_results} we then provide a broad presentation of the scope of the mock spectra library and a preliminary analysis of some of its content. Finally, Section \ref{sec_summary} discusses important caveats, future directions, requests for input from the community, and summarizes the project.


\section{Methods} \label{sec_methods}

\subsection{Ionization States} \label{sec_ions}

In general, most cosmological hydrodynamical simulations do not directly model the ionization state of metals. For any observable absorption line, the ionization fraction of the corresponding species must therefore be derived in post-processing. To do so, we use \textsc{Cloudy} \citep[][v17 and v21]{ferland17}. We include both collisional and photo-ionization processes assuming ionization equilibrium in the presence of a metagalactic background radiation field (i.e. the UVB). For this purpose we prefer to adopt either the same UVB as the hydrodynamical simulation, or a common UVB across simulations for consistency \citep[the 2011 update of][]{fg09}. All processes are always included regardless of which dominates in any particular regime. We use \textsc{Cloudy} in single-zone mode and iterate to equilibrium, accounting for a frequency dependent shielding from the background radiation field (UVB) at high densities \citep[following][]{bird14,rahmati13}.

As gas cells in the simulation are single-zone with no internal structure, it would be inconsistent to assume a multi-zone or more complex geometry in the photoionization calculation. Note that in current cosmological simulations, the size of individual gas cells/particles ranges from $\sim 10$\,pc (in the ISM) to $\sim 10$\,kpc (in the IGM). As we assume gas cells/particles have no internal structure, our final spectra represent only the structure explicitly resolved in the underlying simulation. We run \textsc{Cloudy} in the `constant temperature' mode, with no induced processes \citep[following][]{wiersma09}.

Our fiducial configuration is a 4D parameter space grid in (n$_{\rm H}$,\,T,\,Z,\,z), hydrogen number density, temperature, metallicity, and redshift is produced with the following value ranges: $-7.0 < \log(\rm{n}_{\rm H} [\rm{cm}^{-3}]) < 4.0$ with step size $\Delta$n$_{\rm H}$ = 0.1, $3.0 < \log(\rm{T} [\rm{K}]) < 9.0$ with $\Delta$T = 0.05, $-3.0 < \log(\rm{Z} [\rm{Z}_{\rm sun}]) < 1.0$ with $\Delta$Z = 0.4, and $0 < \rm{z} < 8$ with $\Delta$z = 0.5. Note that the redshift dependence captures the changing intensity and spectral shape of the assumed background radiation field. At each grid point we determine and save the ionization fraction $x_{i,j}$ of the j$^{\rm th}$ ion of species $i$. The total mass (or volume number density) of an ion within a gas cell is then given by this ionization fraction times the mass (or number density) of the parent species.

{\renewcommand{\arraystretch}{1.1}
\begin{table}
    \centering
    \begin{tabular}{p{0.14\linewidth} | p{0.76\linewidth}}
        \hline\hline
        Ion & Transitions [Ang]\\
        \hline\hline
        H I & 1215, 1025, 972, 949, 937, 930, 926, 923, ... \\
        C I & 1561, 1329, 1280, 1277, 1261, 1194, 1193, 1189 \\
        C II & 1334, 1335a, 1335b, 1037, 1036 \\
        C III & 977 \\
        C IV & 1548, 1550 \\
        Ca II & 3969, 3934 \\
        Mg I & 2852, 2026, 1827, 1747 \\
        Mg II & 1239, 1240, 2796, 2803 \\
        Mn II & 2606, 2593, 2576, 1201, 1199, 1197 \\
        N I & 1199, 1134 \\
        N II & 1085, 916 \\
        N III & 990, 764, 685 \\
        N V & 1238, 1242 \\
        Na I & 5897, 5891, 3303 \\
        Ne VIII & 770, 780 \\
        Ni II & 1754, 1751, 1709, 1788, 1741, 1454, 1393 \\
        O I & 1306, 1304, 1302, 1039, 1025 \\
        O III & 834, 703 \\
        O IV & 789, 609 \\
        O VI & 1037, 1031 \\
        O VII & 21, 18, 17a \\
        O VIII & 18a, 18b, 18c \\
        Al I & 3962, 3945, 3093a, 3093b, 3083, 2661, 2653 \\
        Al II & 1670 \\
        Al III & 1854, 1862 \\
        Ar I & 1066, 1048 \\
        Cr II & 2065, 2061, 2055 \\
        Si I & 2515, 2516, 1845, 1847, 1850, 1693, 1696, 1697 \\
        Si II & 1533, 1526, 1309, 1304, 1265, 1264, 1260, 1197, 1194, 1193, 1190 \\
        Si III & 1206 \\
        Si IV & 1393, 1402 \\
        S II & 1259, 1253, 1250 \\
        Ti II & 3383, 3372, 3361, 3349, 3088, 3078, 3075, 3072, 1910 \\
        Ti III & 1298, 1288 \\
        Fe I & 3026, 3021, 2744, 2719, 2541, 2536, 2528, 2525, 2523, 2518, 2490, 2167 \\
        Fe II & 2632, 2631, 2629, 2625, 2622, 2621, 2618, 2614, 2612, 2607, 2600, 2599, 2586, 2414, 2411a, 2411b, 2407, ... \\
        Fe XVII & 15, 13, 12, 11 \\
        Zn I & 2138 \\
        Zn II & 2062, 2025 \\
        \hline
    \end{tabular}
    \caption{$\vphantom{X^{X^{X}}}$ List of the ions, and the transitions (i.e. lines) for each ion, currently considered in SALSA. Each transition gives the vacuum wavelength in Angstroms rounded down. This list is dynamic (see online data repository for availability).}
    \label{table_ions}
\end{table}
}

\begin{figure*}
   \centering
   \includegraphics[width=0.95\textwidth]{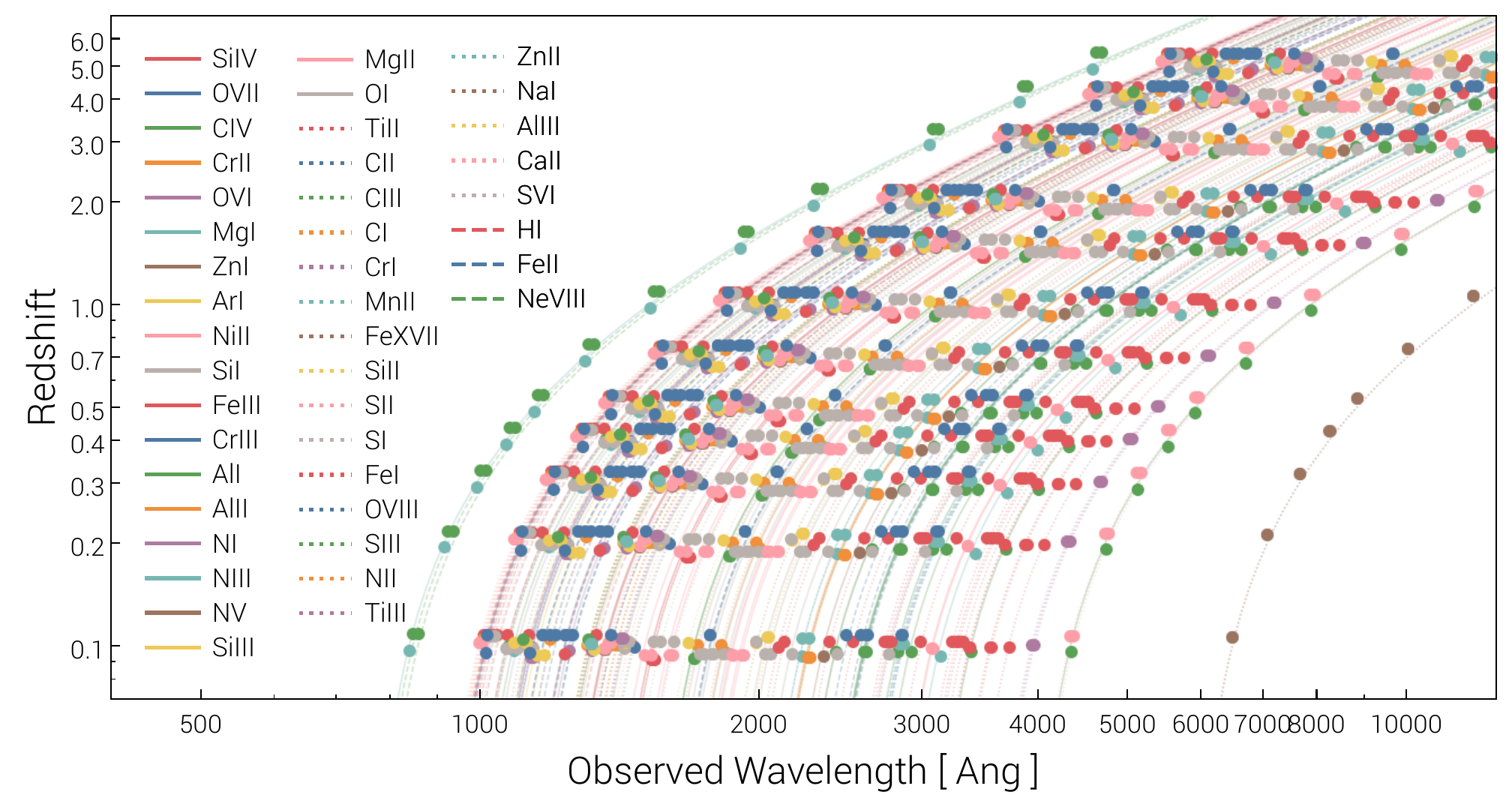}
   \caption{Example of ion and redshift coverage for available mock spectra catalogs. Each marker represents a possible dataset: the combination of a particular ion (colors; upper left legend) and transition and the resulting observed wavelength, at a full TNG snapshot. For each transition we show the evolution of the observed-frame wavelength as a function of redshift (lines). An online, interactive version of information is available in the form of a table, where users can select and download datasets of interest (\url{www.tng-project.org/spectra}).}
   \label{fig:coverage}
\end{figure*}

Some simulations place gas, particular star-forming gas, on an effective equation of state, and this motivates additional care. For example, star-forming gas cells in the TNG model adopt a two-phase pressurization model \citep{springel03}. In this case we explicitly divide each such cell into its cold-phase and hot-phase components, and include both as absorbing media. To do so we calculate the (density dependent) mass fraction of the hot and cold components, as well as the temperature of the hot phase \citep{springel03}. In general, the hot phase contains only a small fraction of mass (that is assumed to occupy the majority of the cell volume): roughly 10\% at the star-formation threshold density of $n_{\rm H} \simeq 0.1 \rm{cm}^{-3}$. This fraction decreases to $\lesssim 1$\% towards higher density. The temperature of the hot phase increases asymptotically from $\sim 10^5$\,K at $n_{\rm H} \simeq 0.1 \rm{cm}^{-3}$ to $\sim 10^8$\,K at high density. As a result, it can host high ionization states of metals, as expected in the hot ISM of galaxies. We caution that, despite this modeling, simulations invoking effective multi-phase ISM models such as \citet{springel03} have limitations within star-forming gas that must be considered with care, and that the absorption arising from this gas is sensitive to sub-grid model choices.

\subsection{Abundances and Dust Depletion} \label{sec_abunddust}

Some galaxy formation simulations explicitly include stellar evolution yields and model the return of mass and metals from stars to nearby gas in the local interstellar medium. Others do not, or include such information only at a subset of outputs. In all cases, only a small list of species are ever tracked, due to computational restrictions. In the case of IllustrisTNG, for example, H, He, C, N, O, Ne, Mg, Si, and Fe are followed \citep[while e.g. Ca, Mn, Na, Al, Ti, and Zn are not;][]{vog13,pillepich18a}, and only available for selected `full' snapshots.

This information, when available, provides a unique view on nucleosynthesis across epochs and environments, as different abundance ratios reflect the distinct progenitor channels of different elements \citep{kobayashi25}. However, even in cases where hydrodynamical simulations model the return of mass and metals from evolving stars, significant uncertainties exist in the stellar yield tables that must be adopted \citep{nomoto13}. Thus, it may sometimes be desirable to assume solar abundance ratios rather than adopt explicitly tracked metal abundances. In general, we make mock spectra for both cases, adopting the solar abundances of \cite{grevesse10} when needed.

In addition, a significant fraction of the total abundance of a given species may be removed from the gas phase and depleted onto dust. Metals locked in dust grains will not contribute to absorption features. Dust depletion is negligible for some species but significant for others \citep[e.g. Fe;][]{jenkins09}. We therefore develop an (optional) simple model to account for the effects of dust depletion. To do so we take the observed dependencies of the dust-to-metal (DTM) ratio on metallicity \citep{peroux20a}, as measured from Damped Lyman-alpha Absorbers \citep[DLAs;][]{decia18}. We fit these data to define a per-species DTM as a function of ISM metallicity [Fe/H], which we then apply on a gas cell-by-cell basis. This empirical model also has uncertainty and is thus optional: absorption spectra with and without dust depletion are available.

\subsection{Sightlines} \label{sec_sightlines}

The creation of synthetic absorption spectra is split into two stages. The first stage generates sightlines by ray-tracing through a simulated volume, and saves the ordered list of intersections between each ray and simulation gas cells. This is a purely geometrical calculation which is agnostic to the physical properties of the gas. The second stage is run for a particular ion. It takes the saved rays as input, computes the necessary gas physical properties, and then deposits absorption profiles in wavelength space in order to construct synthetic spectra.

\subsubsection{The Voronoi Tessellation}

We treat the distribution of simulated gas using a Voronoi tessellation of space. Also known as a Voronoi mesh or Voronoi diagram, in our case this is a spatial decomposition i.e. partitioning of 3D space into a set of $N$ disjoint cells. These Voronoi cells are convex polyhedra, uniquely determined by the specification of a set of $N$ mesh-generating points. Each cell is defined as the region of space closer to its generating point than to any other \citep{dirichlet50}. The natural neighbors of a cell are those that share a common face, i.e. are spatially connected. Voronoi tessellations provide a mesh structure over which geometrical operations such as projection or ray-tracing are well defined and computationally efficient. Voronoi meshes are also naturally and continuously adaptive in spatial resolution. For these reasons, they have been used extensively in the context of radiative transfer (RT) for galaxy formation simulations \citep[e.g][]{camps13,vandenbroucke18,byrohl21}.

The Voronoi tessellation can also be used as a spatial discretization for finite volume type hydrodynamical methods, such as the AREPO moving-mesh code \citep{springel10}. As a result, solving RT-like equations in post-processing directly on the same Voronoi mesh makes the calculation self-consistent with respect to the representation of the underlying gas distribution. This is the case for the IllustrisTNG suite, our primary simulation target, as well as all other simulations run with AREPO including Illustris, Thesan, TNG-Cluster, MillenniumTNG, Auriga, GIBLE, among others. In addition, Lagrangian particle-based hydrodynamics codes including GADGET, GIZMO, and SWIFT and associated simulation projects including EAGLE, FIRE, SIMBA, FLAMINGO, Astrid, and Colibre have gas particle distributions that can be directly represented as Voronoi diagrams, retaining their original density estimates. We therefore develop a ray-tracing approach specifically for Voronoi tessellations as a general and powerful approach.

\subsubsection{Meshless Voronoi Ray-Tracing}

Given a set of $N$ generating points, the corresponding Voronoi mesh can be explicitly constructed. By this we refer to the computation, for each Voronoi cell, of (i) its natural neighbors, i.e. the list of Voronoi cells sharing a face with this cell, and (ii) the geometrical properties of those faces, namely the midpoint between the cell and each neighbor, lying by definition on the face, and the face normal vector, from which the plane equation describing the face can be expressed. Doing so allows tasks such as ray-tracing, projection, and volume rendering to be tackled by walking cell-by-cell through a mesh, identifying subsequent next neighbors via ray-plane intersection tests applied to each face. This is the approach adopted in current Voronoi mesh visualization and RT codes, including in AREPO itself for projection \citep{springel10} and volume rendering \citep{vog13}, in the ArepoVTK visualization package \citep{nelson16}, and for RT in the SKIRT \citep{camps20}, COLT \citep{smith15}, and THOR \citep{byrohl25} codes.

However, for large simulations, explicit construction of the Voronoi mesh is (i) computationally expensive and (ii) prohibitively large to store on-disk. Motivated by these constraints, we develop a new meshless i.e. `mesh-free' algorithm for ray-tracing between points of a Voronoi diagram. The result is exact, yet the Voronoi mesh itself is never computed. The central idea is to ray-trace through a set of generating sites where the intersection point between the current ray position and the next Voronoi face along the ray direction is identified via a fast, (oct)tree-accelerated, spatial bisection search along the ray. The numerical algorithm is discussed in detail in Appendix \ref{sec:voronoi}.

Ray-tracing through a static medium is an embarrassingly parallel problem, and the meshless algorithm parallelizes well, as expected. We use threading and find speed-up factors of 26 (42) for 32 (64) cores, with respect to single-core serial algorithm, corresponding to parallel efficiencies of 81\% (66\%). Additional memory/cache optimizations and better spatial locality could undoubtedly increase this performance. 

We also implement a second level of parallelism via spatial subdivision of the simulation domain. Without scattering, our rays are parallel and have trivial paths. We therefore split the cubic simulation box into $N^2$ elongated cuboids, where the integer $N$ specifies the total number of subdomains. Each cuboid therefore has extent $[L/N+\epsilon, L/N+\epsilon, L]$ where $L$ is the original box-length and $\epsilon$ is a padding factor that is set to comfortably exceed the size of the largest (IGM) gas cells. The work can then be split into $N^2$ jobs, where each is fully independent, can be run simultaneously, and handles only the rays that traverse its subdomain.\footnote{A chunked loading of gas positions lets us compute a spatial mask of snapshot indices, for all future loads. As each job never needs to load the entire snapshot, this strategy also reduces peak memory usage by $\sim N^2$.} With these optimizations, the geometrical ray-tracing step is far cheaper in compute cost than the creation of the actual spectra.

\subsection{Absorption Spectra} \label{sec_mockspec}

At this point we have generated and saved a number of sightlines. Each sightline provides the minimally required information: the ordered list of intersections with simulation gas parcels (i.e. Voronoi cells, identified via their global snapshot indices), and the path-length through each parcel. We then proceed to the second stage, and use this information to create a continuum-normalized absorption spectrum for each sightline, as follows.

For a given instrument we create a master wavelength grid. This spans the instrumental wavelength range, plus some buffer, with a constant spacing i.e. dispersion of $\Delta \lambda = 0.0001$\AA. In some cases, this can lead to as many as $\sim 10^9$ spectral bins, but only one such master grid (per thread) is needed. We opt to explicitly represent a high-resolution master spectrum in memory, instead of using virtual sub-bin deposition, as we discuss below.

For all gas particles/cells in the spatial subdomain, we then load their physical properties: temperature, line of sight velocity $v_{\rm los}$, and density of the particular ion $n$. All transitions of that ion are then processed. For each ray, we compute the cumulative pathlength to each intersected gas parcel, and convert this to a $z_{\rm cosmo}$ using a lookup table of comoving distance versus redshift created around $z_{\rm snap}$ and covering the box extent. For each gas parcel the effective redshift is
\begin{equation}
z_{\rm eff} = (1 + z_{\rm doppler}) (1 + z_{\rm cosmo}) - 1
\end{equation}
where $z_{\rm doppler} = v_{\rm los} / c$. A ray traverses a column density of $N = dl \cdot n$ through a gas parcel with a Doppler parameter $b = (2 k_{\rm B} T / m)^{1/2}$ where $m$ is the mass of the ion. In general we do not add any sub-grid turbulent broadening, such that kinematic broadening occurs (only) due to resolved gas motions. The optical depth as a function of wavelength of each intersected gas parcel is then deposited as a single Voigt absorption profile onto the master spectrum.\footnote{The Voigt profile is the convolution of a Gaussian, that describes broadening due to the velocity distribution of gas, and a Lorentzian, that describes broadening due to the pressure i.e. collisions within gas.} To do so, we iterate to identify a wavelength range wherein the single Voigt profile
\begin{equation}
\tau(\lambda) = \frac{\pi^{1/2} e^2}{m_{\rm e} c} N f \lambda_0 \phi(\lambda)
\end{equation}
drops below an optical depth of $\log{(\tau)} < -4$ at its edges. Here, $f$ and $\lambda_0$ are the oscillator strengths and rest-frame central wavelengths of the transition, respectively. The other fundamental constants have their usual meanings. The dimensionless Voigt profile is calculated as
\begin{equation}
\phi(\lambda) = \rm{Re}\left[\rm{wofz}( u,\alpha )\right] / b
\end{equation}
where $\rm{wofz}$ is the complex-valued Faddeeva function, $u = (\nu - \nu_0) / d\nu$ is the frequency relative to line center ($\nu = c/\lambda$) and $d\nu = b / \lambda_0$. Finally, $\alpha = \gamma / (4 \pi d\nu)$ is the damping parameter, where $\gamma$ is the sum of the transition probabilities (Einstein A coefficients). We use the numerical trick of taking the real part of the Faddeeva integral to efficiently compute the dimensionless Voigt profile \citep{thompson93}.\footnote{We wrap the cython implementation of wofz from scipy to obtain the first 8 bytes, corresponding to the real part (on most architectures), to circumvent the lack of support for complex double return types in numba jitted code.}

The optical depth contributions from all gas parcels are combined, and then converted to flux as $f = \exp{(-\tau)}$. We can then convolve the flux spectrum by the instrumental line spread function (LSF). We use an explicit real-space convolution that allows us to incorporate $\lambda$-dependent LSFs. Finally, we resample the master flux spectrum onto the instrumental wavelength grid.

\subsection{Instrumental and Observational Effects} \label{sec_inst}

{\renewcommand{\arraystretch}{1.3}
\begin{table}
    \centering
    \begin{tabular}{rlc}
        \hline\hline
        Instrument    & Grating(s), Mode(s), etc & LSF \\ \hline
        idealized     & ($\Delta \lambda = 0.01$\AA) & -- \\
        HST/COS       & G130M, G160M, G140L & (1) \\
                      & G185M, G225M, G285M &    \\
                      & G230L               &    \\
        SDSS/BOSS     & composite           & (2) \\
        DESI          & composite           & (2) \\
        4MOST/LRS     & composite           & (2) \\
        4MOST/HRS     & composite, B, G, R  & (2) \\
        WHT/WEAVE     & composite, B, R     & (*) \\
        Subaru/PFS    & B, R-LR, R-HR, NIR  & (3) \\
        Magellan/MIKE & B, R                & (3) \\
        KECK/MOSFIRE  & composite           & (3) \\
        VLT/XSHOOTER  & UVB-05, 10, 16      & (3) \\
                      & VIS-04, 07, 09, 15  &    \\
                      & NIR-04, 06, 12      &    \\
        VLT/UVES      & composite, B, R     & (*) \\
        Gemini/GNIRS  & SXD-R800            & (3) \\
        KECK/HIRES    & B14, B5C3, C4D2, D34, E14 & (3) \\
        KECK/ESI      & 03, 05, 10          & (3) \\
        KECK/LRIS     & B-300, 400, 600     & (3) \\
                      & R-150, 300, 600     &    \\
        ELT/ANDES     & composite           & (*) \\
        \hline
    \end{tabular}
    \caption{$\vphantom{X^{X^{X}}}$ List of spectrographs with instrumental characteristics currently incorporated. For each, different gratings, arms, modes, and so on have different wavelength coverage and LSF properties, that are considered separately. For LSF type, (1) corresponds to a $\lambda$-dependent tabulated LSF, (2) corresponds to a $\lambda$-dependent $R$ value, and (3) corresponds to a constant $R$ value across the full wavelength range. (*) indicates future work for instrument-specific spectra (see online).}
    \label{table_inst}
\end{table}
}

We create an `idealized' set of high-resolution, noise-free spectra. In addition, spectra are created specifically for a given instrumental configuration -- most importantly, to set the wavelength coverage and LSF properties. Table \ref{table_inst} shows the list of setups we currently consider -- these range from past instruments such as the BOSS spectrograph, to current state-of-the-art facilities, to planned, next-generation instruments such as ELT/ANDES \citep{dodorico24}. The spacing of the `wavelength grid' for each instrument is in either linear or log Angstrom, and is chosen to represent common setups i.e. reduction pipelines, such that our synthetic spectra have the same (or similar) numbers of points across the wavelength range. The LSF i.e. spectral resolution can be specified in several ways, in order of decreasing accuracy: $\lambda$-dependent tabulated LSFs (COS), $\lambda$-dependent approximate $R = \Delta \lambda / \lambda$ values (SDSS, DESI, 4MOST), and constant $R$ values across the wavelength range (all other instruments).

For example, for COS-G130M (in the FUV) we adopt $\lambda_{\rm min} = 892$\AA, $\lambda_{\rm max} = 1480$\AA, and $\Delta\lambda = 0.00997$\AA. For COS in particular, we unify the multiple cenwave settings for each grating, that sample the LSF at overlapping wavelengths, by switching when possible in ascending cenwave. By default, we use LP4 for all except G130M at 1055 and 1096 which are LP2. We keep LSFs in pixel coordinates, and do not convert to wavelength space, i.e. we neglect the cenwave dependence of the disptab (wavelength solution), as it is constant for FUV and for NUV we have only a single LSF regardless.

In general, the LSF characteristics of spectrographs are non-trivial to characterize, and various levels of approximation are available. For example, the SDSS LSF characteristics are taken from calibration arc images based on a sample of 100 BOSS plates. Specifically, we take values for spectrograph 1, while spectrograph 2 is worse by $R \sim 100$ at the blue end \citep{smee13}. For DESI, we take SM9 measurements, and have a non-monontonic transition in $R$ between the green and red arms that is only approximate \citep{perruchot20}. For the low-resolution and high-resolution 4MOST spectrographs we take pre-construction estimates, with similar approximations between the arms. In many other cases we have no information beyond a single $R$ value for a given grating \citep{vogt94}. Nonetheless, mimicking the spectral resolution characteristics of the spectrographs under consideration is an important step in forwarding modeling theoretical spectra.

We currently produce only continuum normalized spectra. That is, we assume that the continuum modeling and subtraction of the background source is perfect. In the future, high S/N composite spectra \citep{vandenberk01} or generative theoretical models \citep{sun23} can be used to add (random) continua to the synthetic spectra. Local foregrounds from the Milky Way galaxy or halo \citep{danforth16} can also be  added.

Finally, our synthetic spectra are intrinsically noise free. Some applications will benefit from adding noise. Adding Gaussian random noise with a specific S/N ratio is trivial, and can be requested when loading. However, most spectra have non-trivial noise characteristics. This is particularly evident in the wavelength dependence, where S/N is typically poor at the edges of spectra and the edges of echelle orders. In the future we can incorporate realistic, empirically derived templates of $\lambda$-dependent variance for different instruments.

\subsection{Data Structure and Format} \label{sec_data}

One \textbf{spectra file} exists for a given combination of: (1) simulation, (2) snapshot, (3) sightline configuration, (4) instrument configuration, (5) ion. All considered transitions of this ion are included. The available metadata is (i) the sightline direction, constant for all sightlines, (ii) the starting sightline positions, (iii) the sightline length, constant for all sightlines, and (iii) the wavelength grid, constant for all spectra. Per sightline, and per transition, we make available datasets giving the total equivalent width, column density, and $v_{\rm 90}$ for convenience. In addition, we provide the full optical depth spectra of each transition, and all transitions are combined to form a final dataset of the flux spectra. The latter is just for convenience, and can be reconstructed by summing together the optical depths of all transitions and then computing $f = \exp{(-\tau)}$.\footnote{This (large) file is stored in HDF5, and uses transparent compression to efficiently represent the spectra that are, for most transitions and most sightlines, mainly made up of zeros, i.e. indicating an unabsorbed continuum level.} For example, for a configuration of $2000^2 = 4$ million spectra for SDSS/BOSS, with 4678 wavelength bins, the optical depth and flux datasets have shapes of $[4000000, 4678]$.

Each spectra file is accompanied by a \textbf{sightline file}. This gives the ordered list of gas intersections -- their global snapshot indices, and intersected path-lengths -- for each sightline. This information allows each spectrum to be associated with the individual gas parcels, and thus the \textbf{physical gas properties}, giving rise to the absorption. All properties in the underlying simulation are available, and these typically include temperature ($b$ parameter), density, metallicity (plus chemical composition), position and velocity in 3D space, and so on. In addition, the original \textbf{galaxy and halo catalogs} of the cosmological simulation at each redshift allow absorption features to be correlated with the properties of intervening galaxies and halos. In particular, users can calculate the impact parameter i.e. projected distance from each sightline to all nearby galaxies, recalling that our sightlines are always randomly located.

Figure \ref{fig:coverage} gives an example view of available datasets. In the plane of redshift versus observed wavelength we indicate mock spectra catalogs: each marker corresponds to a specific transition, for a specific instrument, at each available snapshot, as reflected in the horizontal stripes. Similarly, Table \ref{table_ions} lists the ions and corresponding line transitions that we currently consider. Both are subject to additions and updates as time goes on. Up to date documentation, examples of use, and download links are available at \url{www.tng-project.org/spectra}. In addition, an online API is available, such that users can search for, visualize, and download specific spectra of interest.


\begin{figure*}
   \centering
   \includegraphics[width=1.0\textwidth]{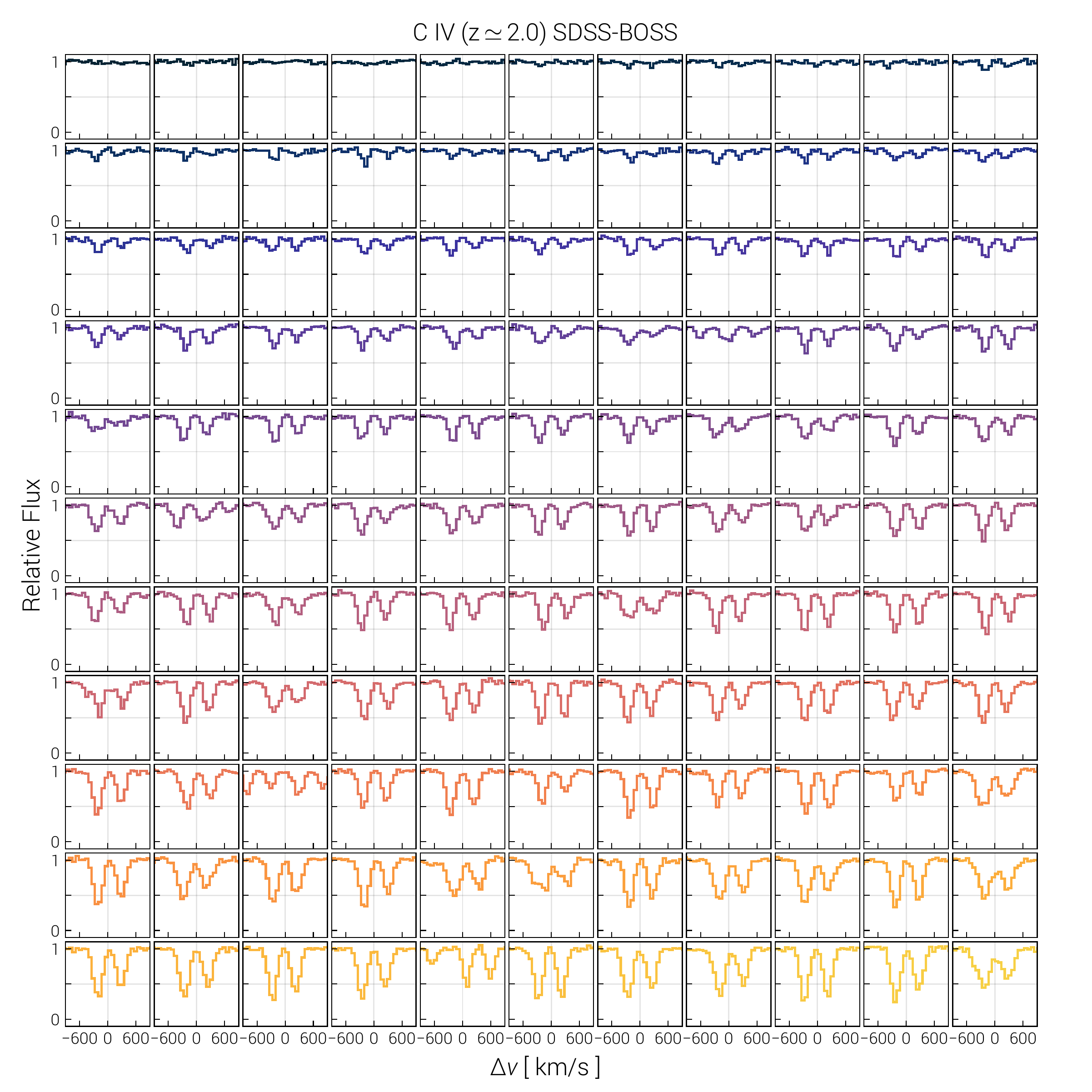}
   \caption{Gallery of mock absorption spectra, focusing on the CIV 1548, 1550\AA\xspace doublet. For a SDSS-BOSS spectral configuration, we randomly select 121 sightlines drawn through the $z=2$ volume of TNG50 that have non-zero CIV equivalent width. These are ordered in ascending EW, from the upper left to the lower right, and show relative i.e. continuum normalized flux in velocity space, centered on the $\tau$-weighted mean wavelength. They pass through a variety of CIV bearing gas structures, from the IGM, to the CGM and ISM of galaxies, all of which imprint a diverse variety of observable absorption signatures in these synthetic spectra.}
   \label{fig:spectra_gallery}
\end{figure*}

\section{The SALSA Project} \label{sec_results}

To showcase the breadth of available data products and to inspire diverse use cases, we highlight a few examples of synthetic absorption spectra and their applications in SALSA.\footnote{No relation to the SALSA software package of the same name from \citet{boyd20}.} In Figure \ref{fig:spectra_gallery} we begin with a gallery of $\sim 120$ CIV 1548, 1550\AA\xspace doublet features, taken from mock SDSS-BOSS spectra of TNG50 at $z=2$. These are evenly sampled in equivalent width, from $0.1$\AA\xspace (black, upper left) to $5$\AA\xspace (yellow, lower right). The weakest absorption features disappear in the $\rm {SNR} = 50$ spectra, while the strongest reach relative fluxes of $\sim 0.5$ and show a variety of complex structure that arises as each sightline passes through spatially and kinematically resolved gas structures in this cosmological hydrodynamical simulation.

In particular, while some spectra could arise from `single-cloud' absorption, many show more complex signatures, even at this relatively low spectral resolution. These trace absorbing gas that has resolved line-of-sight density and kinematic structure in the hydrodynamical simulation, and provide non-trivial examples of absorption features.

\begin{figure*}
   \centering
   \includegraphics[width=0.46\textwidth]{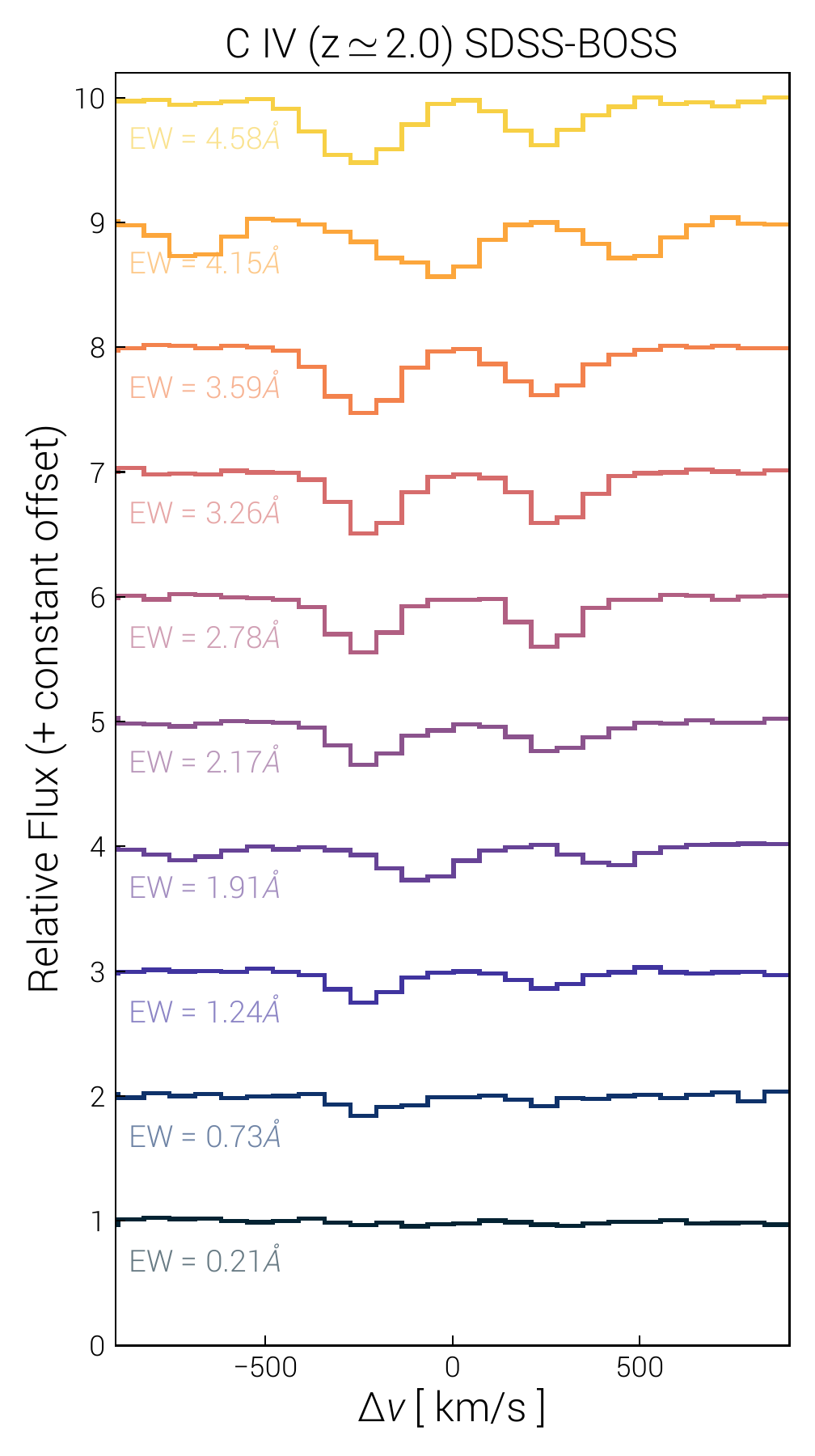}
   \includegraphics[width=0.46\textwidth]{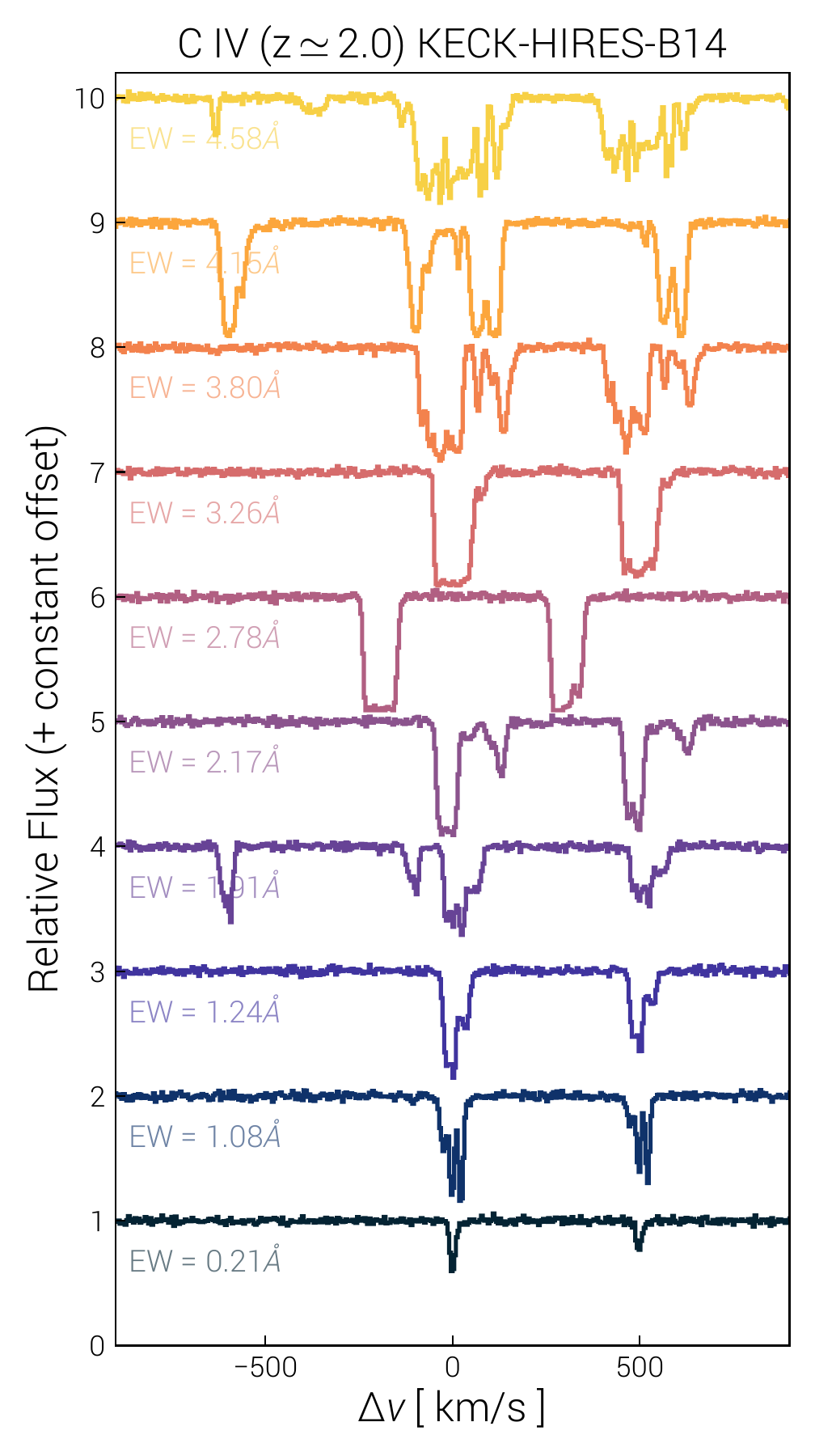}
   \caption{Ten mock absorption spectra from TNG50 at $z=2$, selected randomly such that they are evenly spaced in total CIV equivalent width from $\sim 0.3$\AA\xspace to $\sim 4.6$\AA\xspace. Each shows relative i.e. continuum normalized flux in velocity space, centered on the $\tau$-weighted mean wavelength. Left versus right panels compare the same sightlines, and same absorption features, as observed with two different instruments: SDSS-BOSS (left) and KECK-HIRES (right). Complex, multi-component structure is evident at sufficiently high velocity resolution.}
   \label{fig:spectra_CIV}
\end{figure*}

Figure \ref{fig:spectra_CIV} shows a more detailed view of ten randomly selected CIV spectra, also from TNG50 at $z=2$. For the same ten sightlines, that intersect the same gas structures, it contrasts the resulting mock spectra generated for SDSS-BOSS characteristics (left) versus those obtained for the KECK-HIRES instrument (right). While the BOSS LSF has a FWHM of $\sim 3$\AA\xspace and so a velocity resolution of $\sim 70 \rm{km s^{-1}}$, the HIRES spectrograph has $R \simeq 67000$ that gives it a $\sim 25-50$ times better spectral resolution. As a result, CIV absorption features that appear relatively simple and possibly consistent with single cell (i.e. single cloud) absorption are in fact comprised of complex sub-component structure. This is particularly true for the stronger, high EW absorption sightlines. In these cases, up to $\sim 10$ discrete absorbers are visible in each component of the doublet.

This is a case where the ground truth information from the hydrodynamical simulation holds particular added value. The line-of-sight integral passes through discrete i.e. spatially disconnected regions of high CIV number density, and each region is resolved by multiple gas cells. These discrete structures can be identified, counted, and correlated with the velocity range over which each contributes optical depth within the observable spectrum. As one application, this enables a forward-modeling based approach (i.e. simulation-based inference) to extract physical information -- such as the number of discrete clouds, and the underlying physical properties of their gas -- from observable spectra.

\begin{figure}
   \centering
   \includegraphics[width=0.46\textwidth]{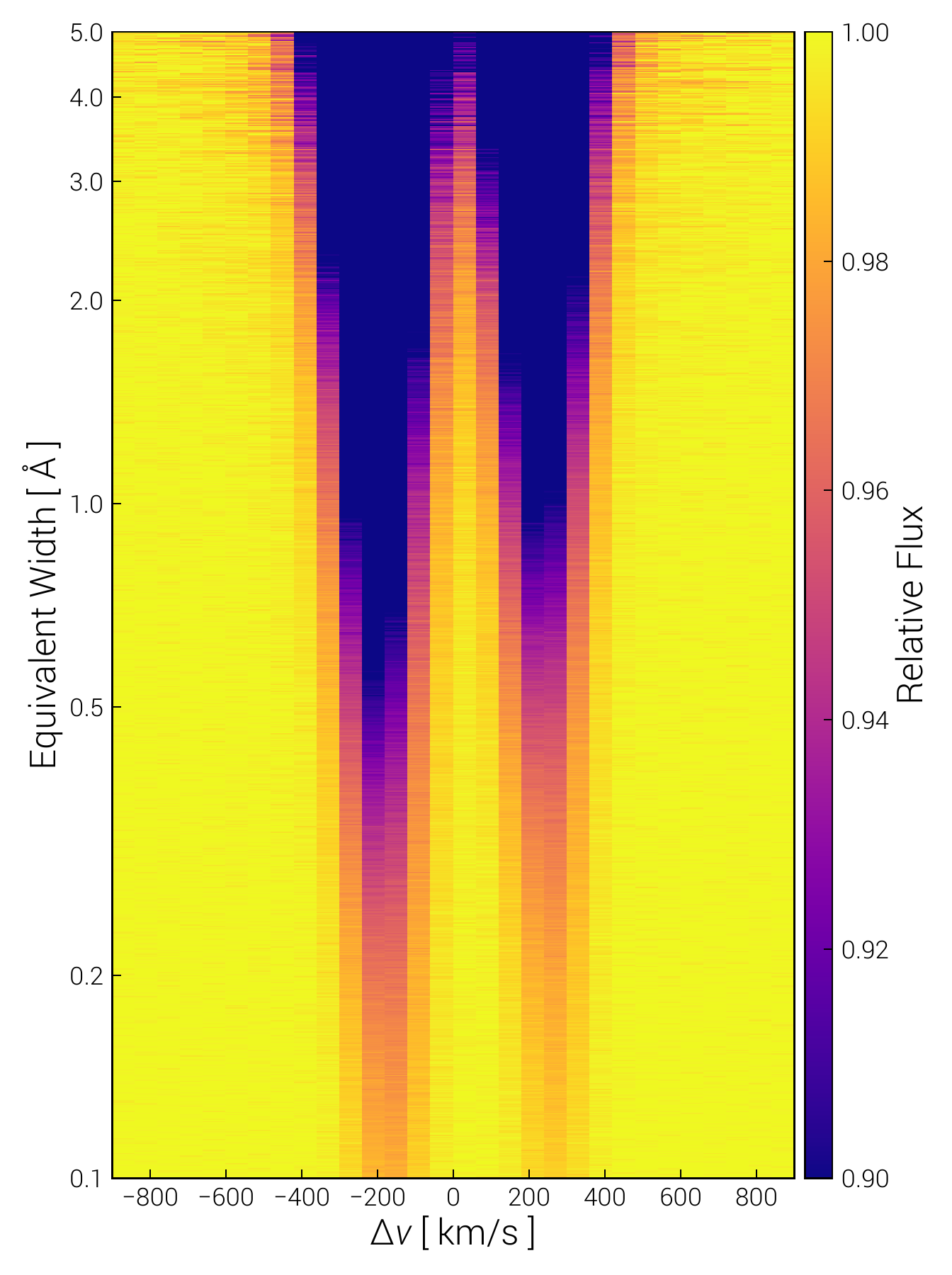}
   \caption{2D stacked gallery of absorption spectra focusing on the CIV 1548+1550 doublet, where color shows relative flux. From the 1M sample of TNG50 at $z=2$ we include 182,744 spectra with $\rm{EW_{CIV}} > 0.1$\AA. These are sorted in descending equivalent width, from top to bottom, where each row shows an individual spectrum. The large statistics of the mock spectra catalogs enables use cases that require abundant statistics and good sampling of the diversity of gas structures giving rise to absorption throughout the diffuse Universe.}
   \label{fig:spectra_2d}
\end{figure}

One pre-requisite for many use cases is ample statistics. Figure \ref{fig:spectra_2d} visualizes a single sample of 182,744 spectra (out of one million) with $\rm{EW_{CIV}} > 0.1$\AA\xspace, available from a particular sightline configuration drawn through the $z=2$ snapshot of TNG50. These are sorted, from top to bottom, in order of decreasing EW, and shown as a two-dimensional stack as a function of wavelength. The intrinsic doublet ratio is clear at low EW, which gives way to saturation and eventually blending of the two components at high EW. Absorption structure beyond the pure double-Voigt profile is also increasingly apparent at large EWs, due to complex underlying gas distributions along each sightline.

A significant strength of mock absorption spectra generated from cosmological hydrodynamical simulations is the realization of non-trivial absorption profiles. These arise naturally from gas structures that are morphologically and kinematically inhomogeneous, i.e. clouds that have internal density and velocity structure, as well as the line-of-sight superposition and close proximity of different absorbing gas structures. The relationship between equivalent width and column density reveals some of this complexity.

\begin{figure}[t]
   \centering
   \includegraphics[width=0.46\textwidth]{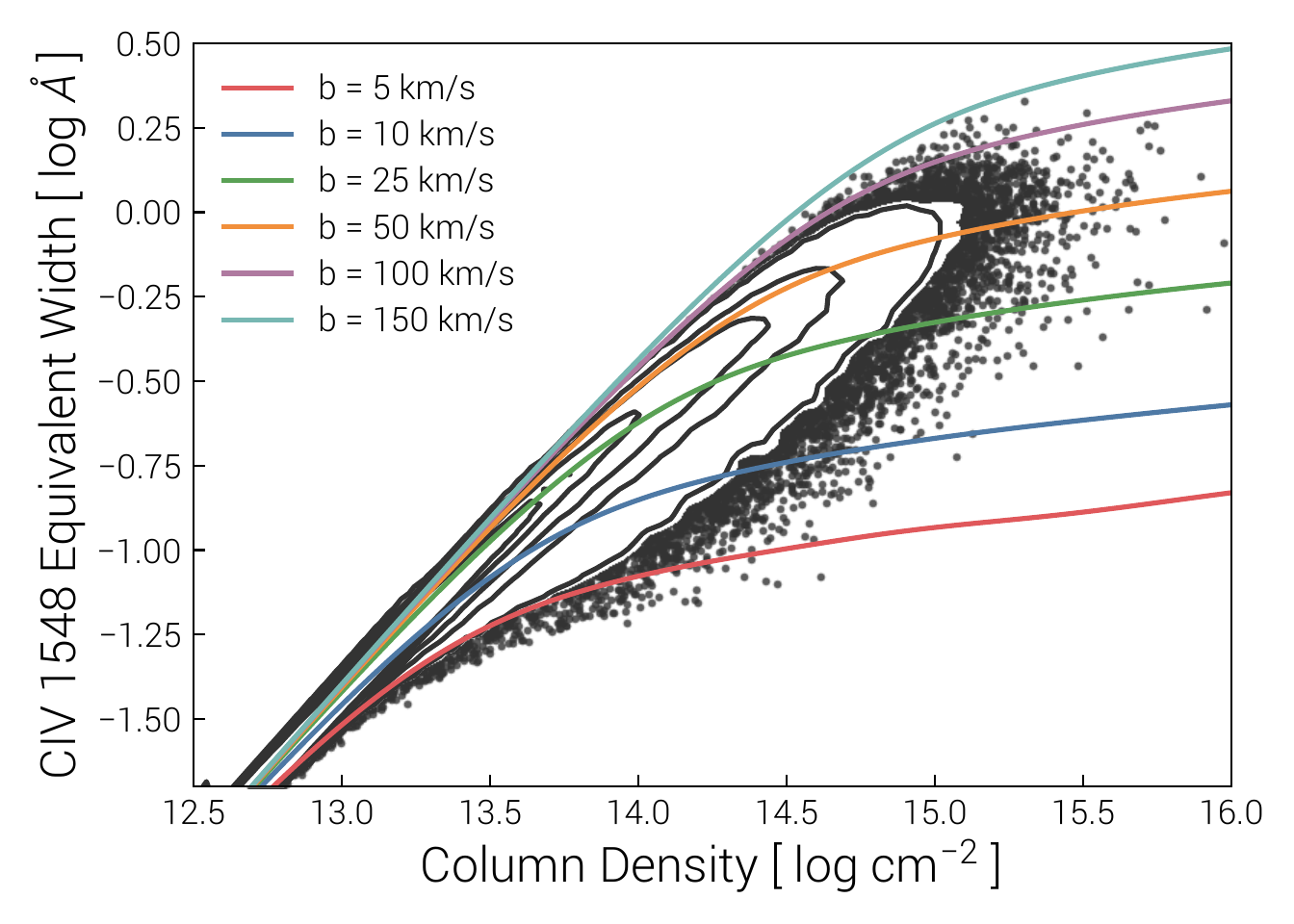}
   \includegraphics[width=0.46\textwidth]{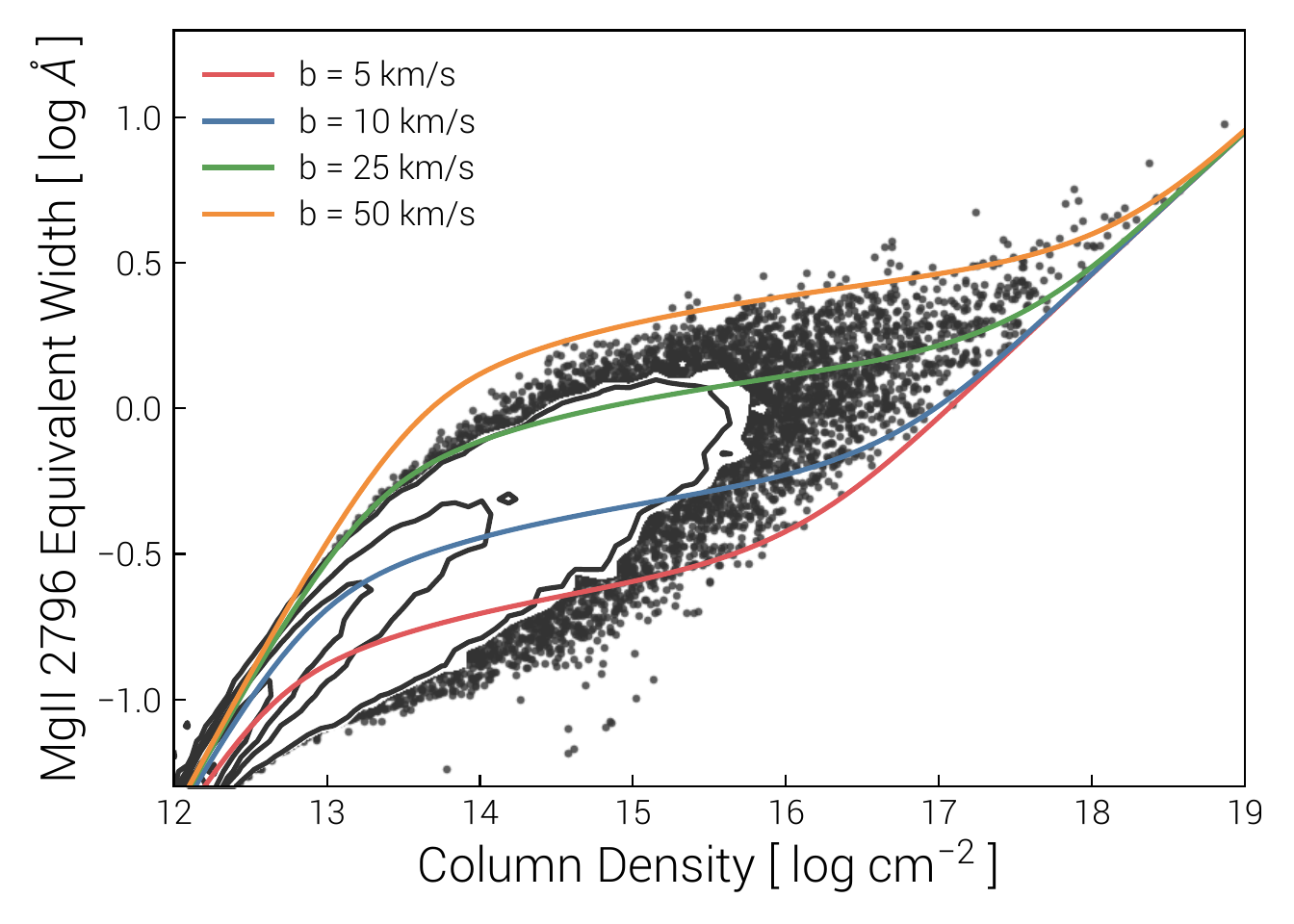}
   \caption{\textbf{Top:} CIV 1548 equivalent width as a function of CIV column density. The six colored lines show theoretical curve of growth calculations for increasing Doppler parameters $b$, from $5\,\rm{km\,s^{-1}}$ to $150\,\rm{km\,s^{-1}}$. The black distribution in the background, visualized with contours plus outliers as individual points, shows the simulation result for TNG50 at $z=2$ derived directly from one million SDSS-BOSS mock spectra. \textbf{Bottom:} MgII 2796 equivalent width as a function of column density.}
   \label{fig:cog}
\end{figure}

In particular, Figure \ref{fig:cog} shows the theoretical curve of growth (CoG) for CIV 1548 (top panel) and MgII 2796 (bottom panel), for different Doppler $b$ parameters ranging from $5$\,km/s to $150$\,km/s (colored lines). Behind, the black contours and scattered points show this relationship for CIV absorption from SDSS-BOSS mock spectra. The EW is integrated from the noise-free spectra, while the column density is the direct integral of the underlying density field along the same sightlines. 

First, we see significant scatter in EW at fixed column. This suggests that CIV absorption can arise in gas across a broad range of temperatures, potentially spanning photoionized and collisionally ionized regimes \citep{davies23b}. It could also reflect significant non-thermal contributions to line broadening, due to internally resolved motions including turbulence in CIV absorption gas \citep{kumar24}, as well as variation of $b$ itself. On the other hand, the optically thick case of $\tau \gtrsim 1$ loses the additive relation between EW and column density if more than one absorber is present in the same integrated velocity range. The broad distribution from the simulation therefore also hints towards the common multiplicity of absorbers and the difficulty of naively inverting (total, i.e. non-component wise) EW to obtain column density \citep{dodorico10}. Nonetheless, we see that the simulation ridge follows most closely the $b \sim 25-50$\,km/s lines \citep[green and orange; see e.g.][]{banerjee23}, depending on column density.

The lower panel of Figure \ref{fig:cog} shows the same relationship between EW and column density for the MgII 2796 transition. In this case the values from TNG50 at $z=2$ span $b$ parameters from $\sim 5$\,km/s to $\sim 50$\,km/s. Particularly for higher column density absorbers, high multiplicity i.e. the number of distinct components can be large \citep{churchill20}.

\begin{figure*}
   \centering
   \includegraphics[width=0.49\textwidth]{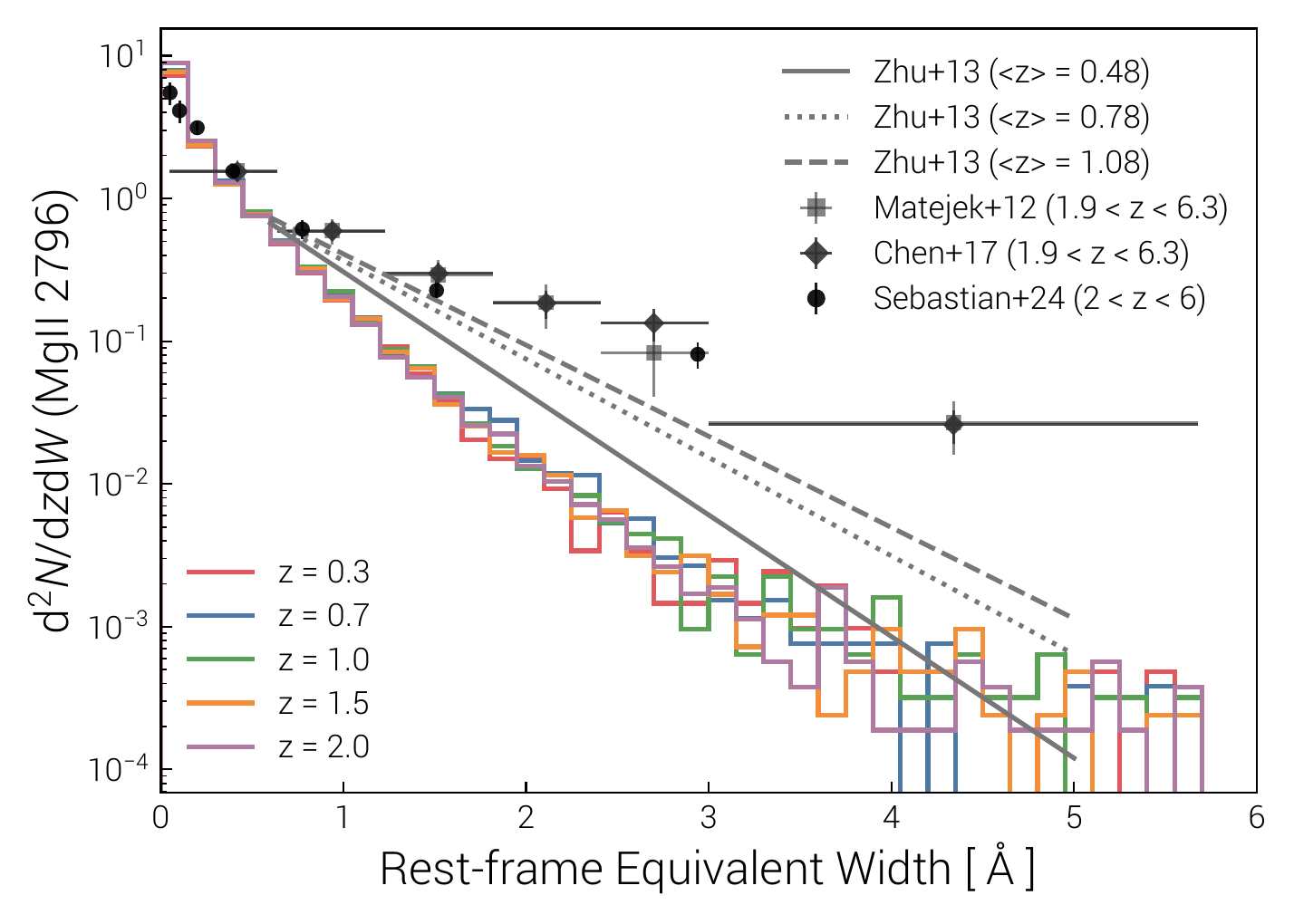}
   \includegraphics[width=0.49\textwidth]{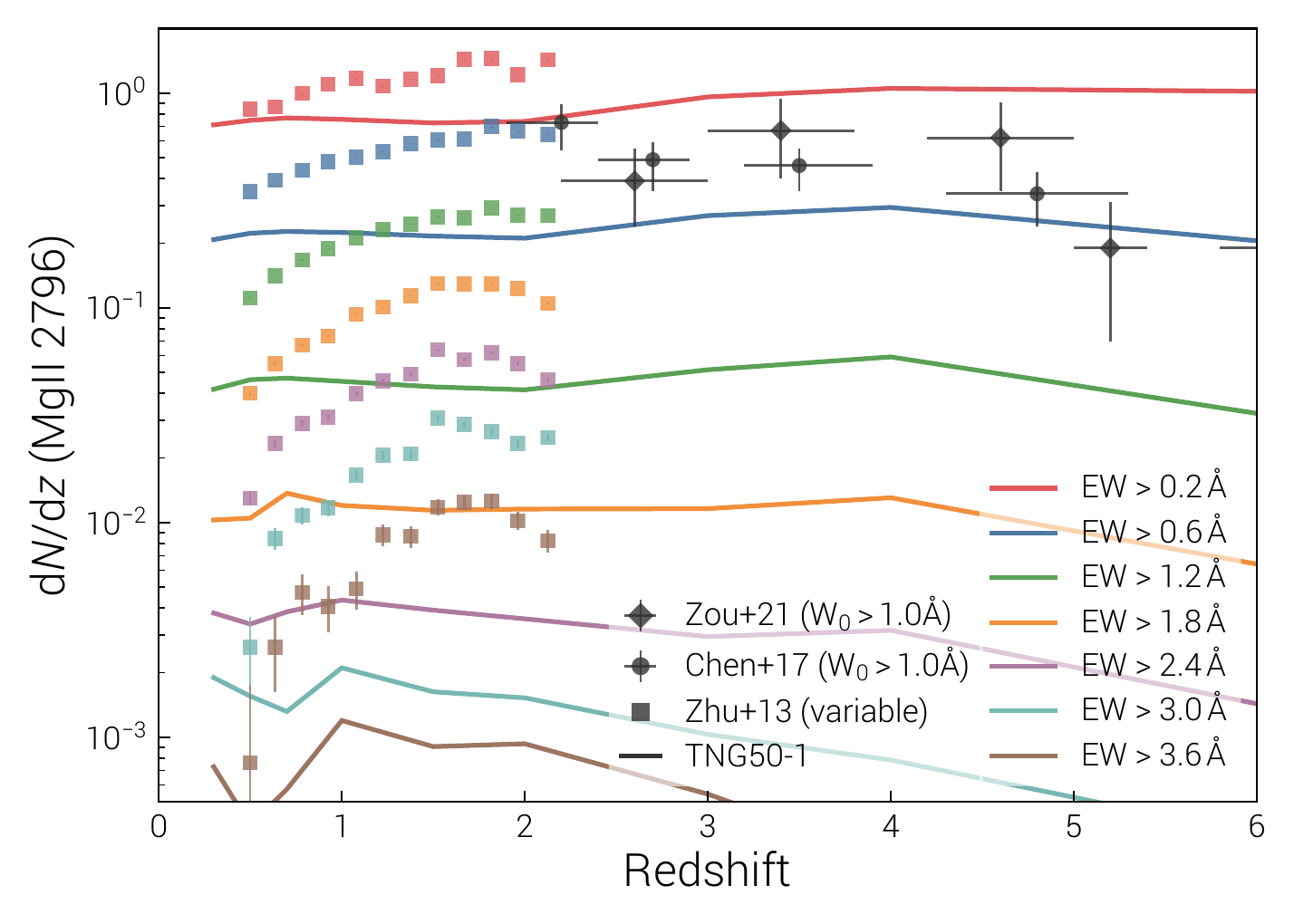}
   \includegraphics[width=0.49\textwidth]{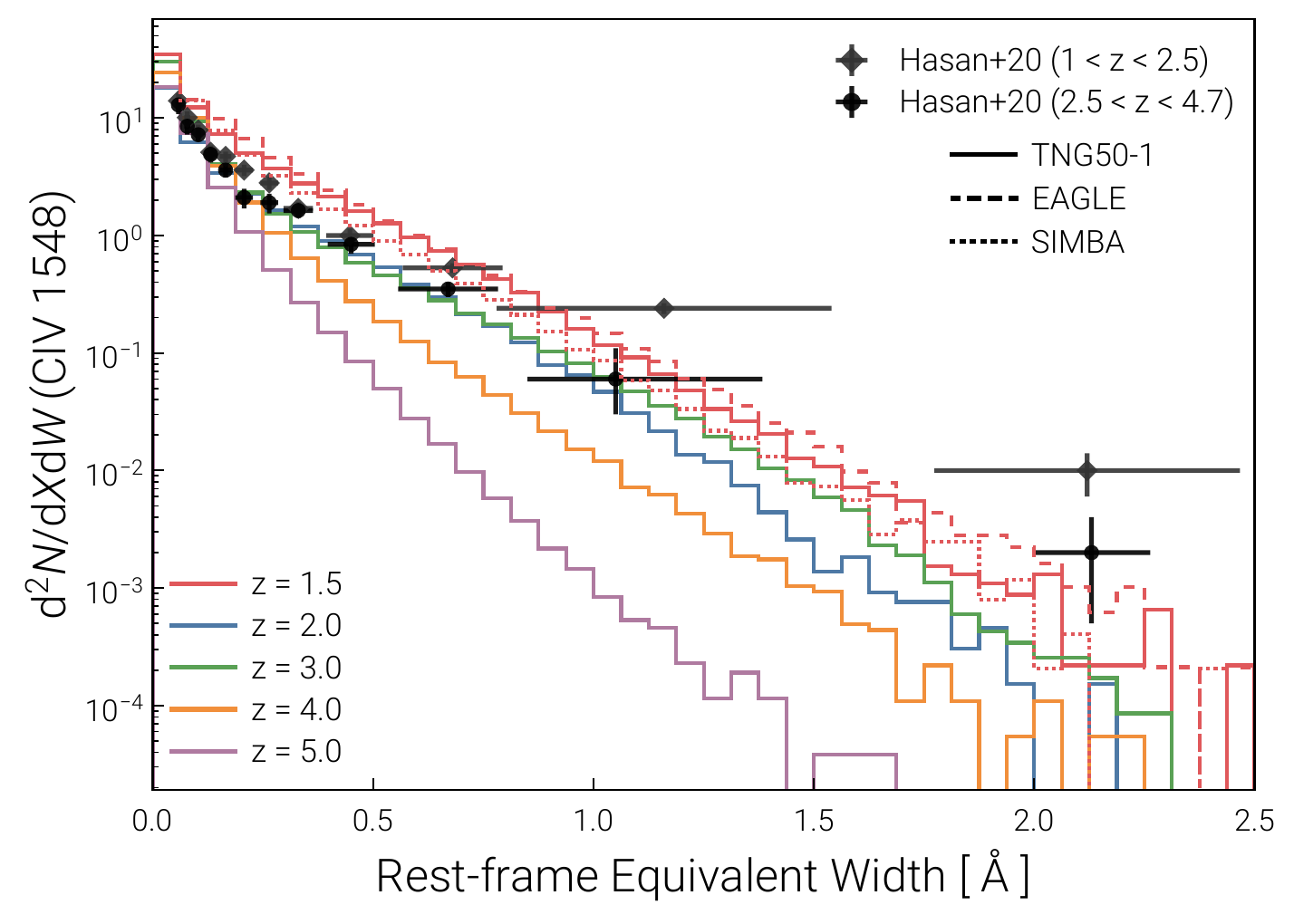}
   \includegraphics[width=0.49\textwidth]{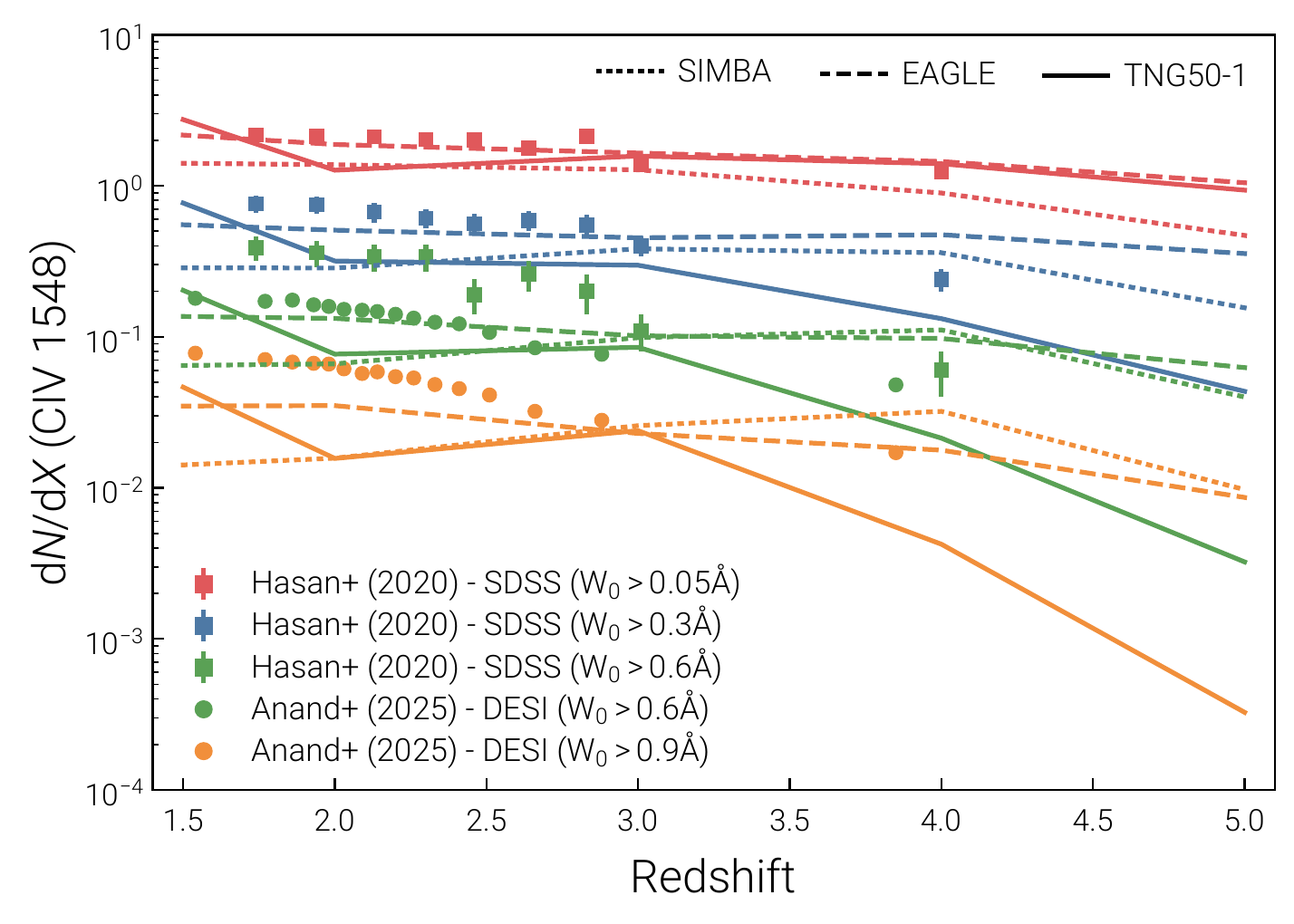}
   \caption{Example of statistics drawn from blind quasar absorption surveys: distribution functions of absorbers as a function of equivalent width (left column), and the redshift evolution of absorber incidence (right column). Broadly tracing cooler versus warm phases, we compare MgII 2796 (top panels) to CIV 1548 (bottom panels). Results are drawn from mock SDSS-BOSS spectra derived from TNG50 across a range of redshifts, as well as EAGLE and SIMBA with DESI mocks in select cases, where we always treat the $10^6$ sightlines at each simulation snapshot redshift as random `on the sky'. This enables comparison to statistics derived from observational surveys that obtain large samples of quasar absorption spectra but do not try to correlate or connect individual absorbers to individual foreground galaxies.}
   \label{fig:blind_stats}
\end{figure*}

The statistics of absorption in `blind' surveys provide an unbiased view of the physical state of diffuse gas as well as its cosmic evolution. Many statistics are measured to quantify the incidence of absorption, typically as a function of strength and redshift. Figure \ref{fig:blind_stats} gives several illustrative examples, focusing on MgII (top row) and CIV (bottom row), with theoretical expectations from TNG50, EAGLE, and SIMBA across $0.3 < z < 5$ (colored lines) contrasted against a variety of observational data (points with errors).

The upper-left panel shows the equivalent width distribution function of MgII 2796, defined as the number of absorbers per unit (rest-frame) equivalent width and redshift. Here we compare to observational results that stack across $2 \lesssim z \lesssim 6$ \citep{matejek12,chen17,sebastian24}. We also compare to low-redshift results from SDSS \citep{zhu13}. The lower-left panel shows the EW distributions for CIV 1548, tracing warmer gas phases, at intermediate to high redshifts. Here we also include observational results on the equivalent width distribution function of CIV \citep{dodorico13,hasan20}. In TNG50, for instance, there is a strong difference in the redshift evolution of these two tracers, which is negligible in the case of MgII but significant for CIV. While the CIV results are reasonable at all redshifts, the MgII EW distribution falls below the high-redshift data. We speculate that the lack of ultra-strong MgII absorbers \citep{udhwani25,fernandez25} may reflect either resolution or physical model limitations, i.e. TNG50 may not resolve the full abundance of MgII clouds, or the presence of MgII in the star-forming ISM of galaxies.

The upper-right panel shows the redshift evolution of MgII 2796, as quantified by the redshift path density of absorbers $\rm{d}N/\rm{d}z$. This is defined as the number of absorption features of a given equivalent width or greater per unit of redshift \citep{sargent88}. We compare results from mock TNG50 SDSS-BOSS spectra against observational data \citep{zhu13}. For $z=2$ we also compare mock X-SHOOTER spectra from TNG50 against observations \citep{chen17,codoreanu17,zou21} At low redshift, the dependence on EW threshold is comparable between the simulation and observations. The redshift trends evident in TNG50 are nearly flat, and thus weaker than implied by some data at $z<2$, although other recent observational analyses are more consistent with a slow evolution of MgII incidence since reionization, depending on EW \citep{lopez25,churchill25}. At high redshift, the incidence of strong MgII absorbers appears weaker than seen in data \citep[see also][]{defelippis21,cherrey25}.

The lower-right panel shows the redshift evolution of CIV 1548, given in terms of the comoving path density of absorbers $\rm{d}N/\rm{d}X$ above an equivalent width threshold $W > W_0$. We again compare SDSS-BOSS-like TNG50 spectra to observational data from SDSS \citep{hasan20} and DESI \citep{anand25}. Observations are shown as symbols with errorbars, while the same four EW thresholds (different colors) are shown for TNG50 with solid lines. In this case, trends in both EW threshold and redshift are qualitatively similar, including the dependence on absorption strength, with a possible slight underabundance of the strongest CIV absorbers. We showcase the cross-simulation comparison utility of SALSA by also including results from EAGLE (dashed) and SIMBA (dotted) as well. Different redshift trends highlight sensitivity to differences in physical modeling. Here in particular the high-redshift evolution of CIV is a good example of sensitivity to radiative processes related to the UVB \citep{huscher25}, while divergence for low-redshift strong absorbers (orange) may reflect AGN feedback-driven heating.

Such comparisons of blind absorption statistics provide informative constraints on hydrodynamical simulations and their ability to produce different gas phases as a result of various astrophysical processes, from cooling to supernova and AGN feedback, i.e. across the global baryon cycle.

\begin{figure*}
   \centering
   \includegraphics[width=0.33\textwidth]{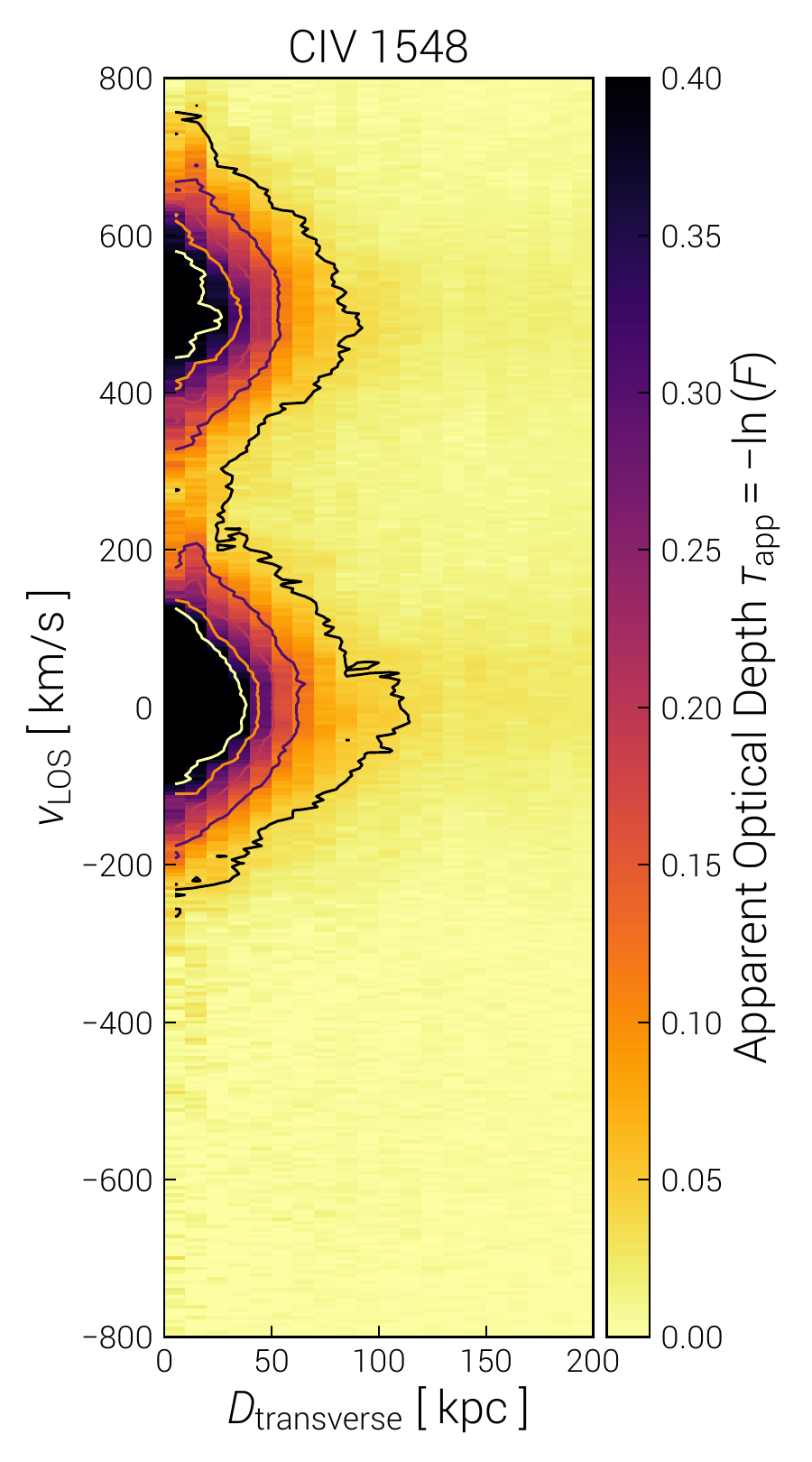}
   \includegraphics[width=0.33\textwidth]{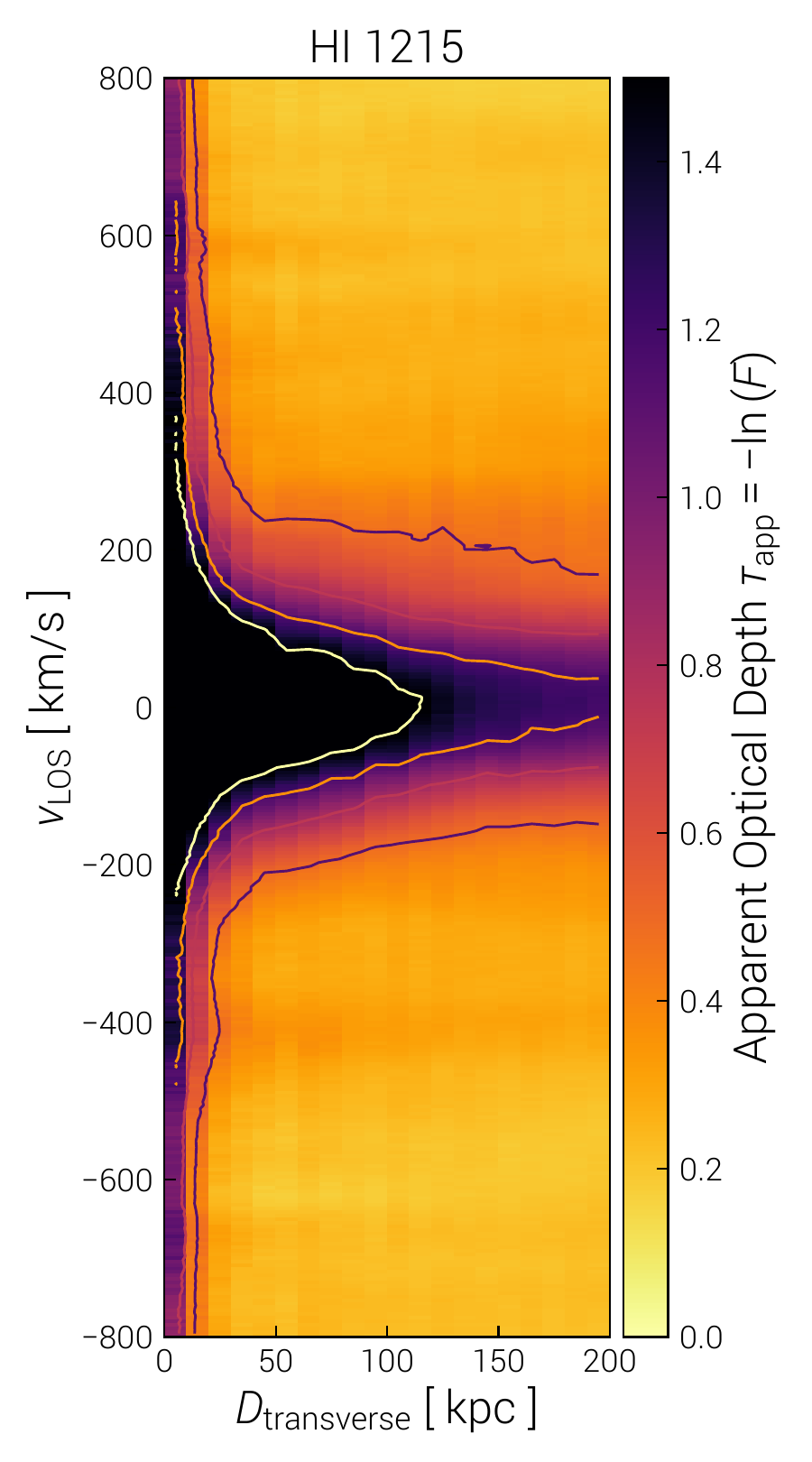}
   \includegraphics[width=0.33\textwidth]{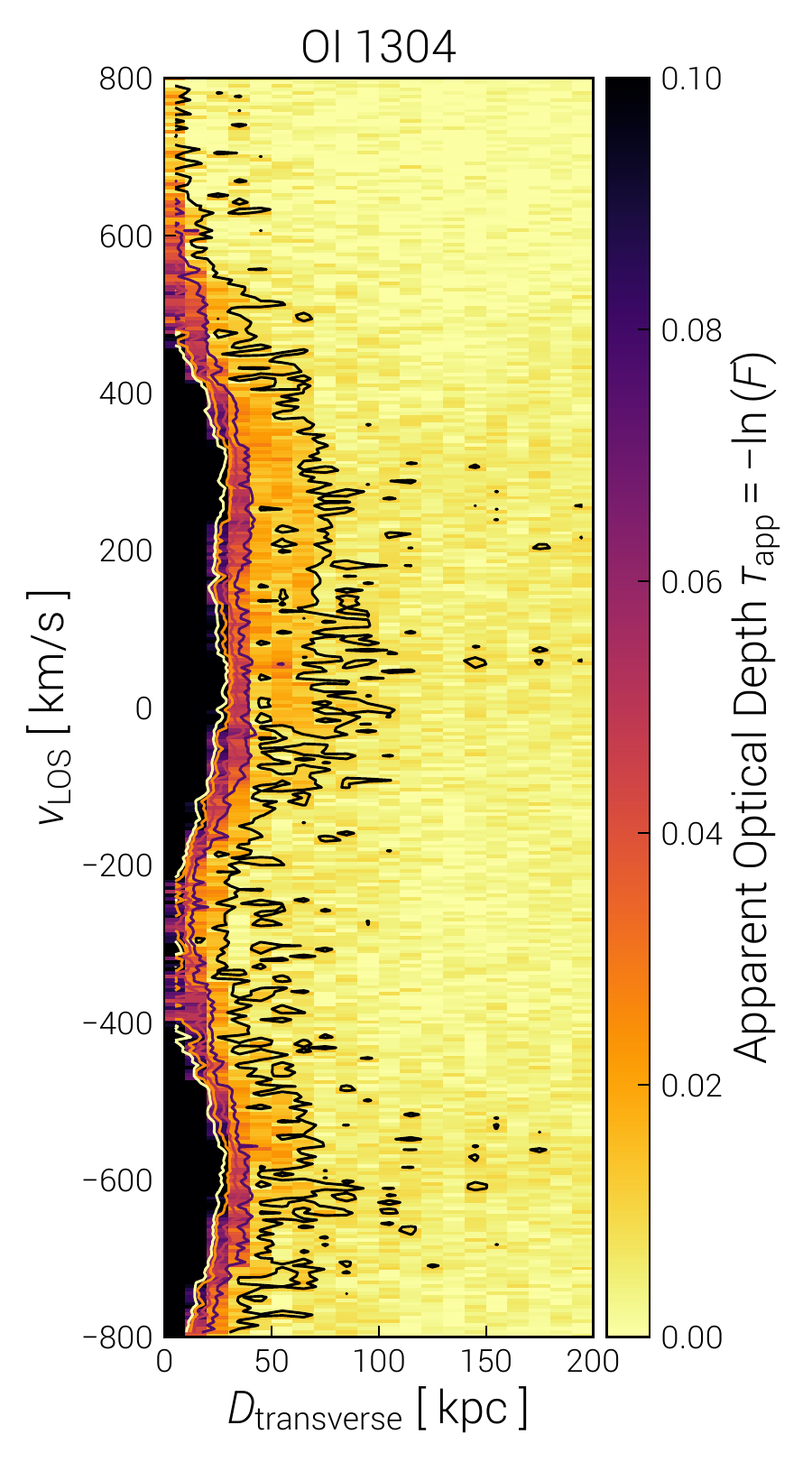}
   \caption{Example of galaxy-centric absorption spectra analysis: two-dimensional apparent optical depth maps. We stack all sightlines that pass within a projected (transverse) distance of $M_\star \simeq 10^{10}$\msun galaxies at $z=2$ in TNG50-1. We then mean stack the KECK-HIRES normalized flux spectra, re-centering each at the expected redshifted emission wavelength of the transition, given the galaxy redshift. We assume each original (single) spectrum has $\rm{SNR} = 10$. For comparison, we stack on CIV 1548 (left), HI 1215 (middle), and OI 1304 (right). In the first case, the CIV 1550 doublet component becomes visible at $v_{\rm LOS} \simeq 500 \rm{km s^{-1}}$, while the HI feature is isolated across this velocity range, and the OI 1304 shows broad structure at small distances with contributions from its two neighboring lines. This tomography reveals extended, halo-scale absorption, with LOS smearing due to the local Hubble flow.}
   \label{fig:tau_map2d}
\end{figure*}

We highlight another important and common use case for absorption spectra: galaxy-centric analyses. Figure \ref{fig:tau_map2d} shows a mock survey experiment of background-foreground pairs that are used to synthesize 2D apparent optical depth maps. In particular, we first identify a foreground galaxy sample using TNG50-1 at $z=2$, selecting solely on galaxy stellar mass within $9.9 < \log{(M_\star / \rm{M}_\odot)} < 10.0$ to obtain about $\sim 100$ galaxies. From a catalog of one million random $\rm{SNR} = 10$ sightlines from this same simulation volume, we identify those that pass within 250\,pkpc in projected i.e. transverse distance from any of these galaxies. This results in $\sim 10^4$ pairs, which we stack on the expected wavelengths of CIV 1548, HI 1215, and OI 1304, given the galaxy redshifts and peculiar velocities. The resulting mean normalized flux spectra are combined in 10\,pkpc bins of impact parameter, and then the apparent optical depth is taken as $\tau = -\ln{(F)}$.

The tomography enabled by these galaxy-centered 2D optical depth maps reveals that significant absorption extends out to transverse distances of $\sim 50-200$\,pkpc, somewhat larger than the virial radii of these halo masses at $z=2$. In the line-of-sight direction, absorption is present at halo-scales out to roughly $\pm 250 \,\rm{km s^{-1}}$, corresponding to $\sim 1$\,pMpc if due to the Hubble flow, as suggested by elongation along the LOS. Such an analysis is inspired by, and qualitatively similar to, findings from 2D apparent optical depth maps from KBSS \citep[e.g.][]{chen21,turner14a,prusinski25}. This demonstrates the powerful utility of combining our mock absorption spectra catalogs with coincident galaxies. In particular, with the galaxy catalogs that are also available from each cosmological hydrodynamical simulation. The detailed (observable) properties of galaxies available makes it possible to replicate, assess, and/or optimize specific galaxy selection functions, and to identify expectations for diffuse gas absorption around galaxy samples with different selection functions and/or intrinsic physical properties, from stellar mass, to star formation activity, to large-scale environment, and so on \citep[e.g.][]{weng24}.

\begin{figure*}
   \centering
   \includegraphics[width=1.0\textwidth]{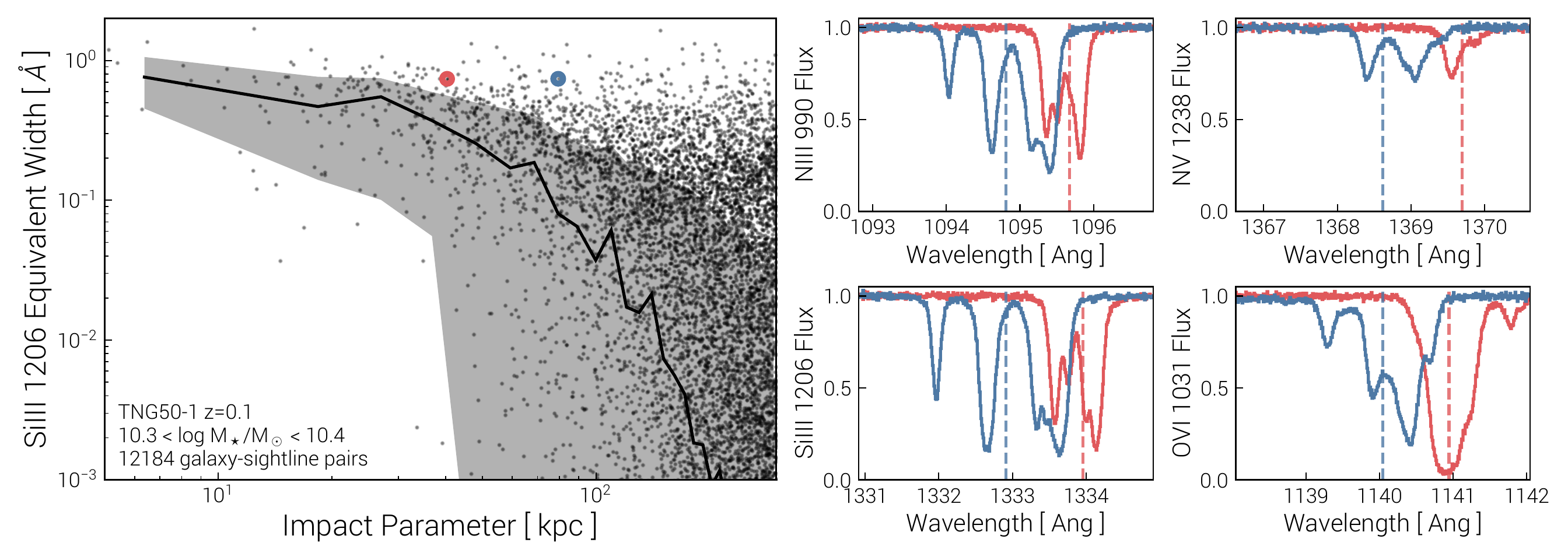}
   \caption{Galaxy-centric absorption spectra analysis: trend of absorption equivalent width versus impact parameter (main panel). We include all sightlines that pass within a projected (transverse) distance of $M_\star = 10^{10.3-10.4}$\msun galaxies at $z=0.1$ from TNG50. There are 95 such galaxies, and using a sample of $10^6$ spectra there are 12184 galaxy-sightline pairs with a projected distance of $b < 300$ pkpc. The black line and shaded band show the running median and $16-84$ percentile band, respectively. Two random high-EW sightlines are highlighted in red and blue, and the corresponding COS-G130M spectra (SNR=100) are shown for four transitions: NIII 990, NV 1238, SiIII 1206, and OVI 1031 (right panels). In each case the systemic redshift of the galaxies are shown with the vertical dashed lines.}
   \label{fig:EW_profile}
\end{figure*}

Figure \ref{fig:EW_profile} shows an example of another common survey technique in CGM surveys: galaxy-centric profiles of absorption strength. We take galaxies from TNG50-1 at $z=0.1$ within a narrow stellar mass range of $10.3 < \log M_\star / \rm{M}_\odot < 10.4$, giving a sample of 95 systems. With a sample of $10^6$ sightlines through this snapshot, our `virtual survey' has $\sim 12,000$ galaxy-sightline pairs with impact parameter $b < 300$ kpc. We then construct mock COS-G130M spectra for a number of commonly observed low redshift transitions: NIII 990, SiIII 1206, NV 1238, and OVI 1031. Note that other common tracers such as HI, NII, CII, and SiII are also available.

The main panel shows SiIII 1206 equivalent width as a function of projected impact parameter. Each galaxy-absorption sightline pair is given by a single dot, while the running median and $16-84$ percentile band is shown with the black line and shaded band, respectively. The well known anti-correlation of cool ion absorption with impact parameter is clear \citep{nateghi24}. Note that in a real observation, the vast majority of the low EW values would be upper limits.\footnote{At large impact parameter, our median relation drops below the visible points due to many weak absorbers with $\rm{EW} < 10^{-3}$\AA\xspace that are not visible.}

To explore the structure of absorption as resolved by COS, we focus on two random sightlines, highlighted by the red and blue circles in the main panel. For these two systems, we show the G130M continuum-normalized spectra (four smaller panels). The vertical dashed lines indicated the systemic redshift of each galaxy, i.e. the wavelength where the transition should be centered if at the same effective redshift as the (nearest) galaxy. We contrast NIII 990, SiIII 1206, NV 1238, and OVI 1031. These trace different phases of the gas, show different component structure, and have different kinematic associations with the galaxy. These $\sim 0.8$\AA\xspace EW (SiIII 1206) examples show significant multi-component structure, with clear associations between the NIII and SiIII features. However, the warmer gas tracers do not necessarily trace the same structure, and warmer components do not correspond exactly to cooler tracers. The detailed structure of OVI and relationship between OVI and NV is a good example of where inclusion of non-equilibrium effects in the future may be important. Revealing the links between multi-phase gas components is one use case of the synthetic spectral libraries.

Our main data product are sightlines that intersect the volume of a simulation box once. As a result, they are typically `short' in terms of cosmological distance, particularly in comparison to observational spectra of high-redshift background targets with significant foreground path-length. One can derive cosmological lightcones from the discrete snapshot outputs of simulations such as TNG \citep[e.g.][]{shreeram25}, and sightlines with corresponding absorption spectra can be computed directly on such lightcones. In addition, some recent large-volume simulations, namely FLAMINGO and MTNG, have begun to write on-the-fly lightcone particle outputs, with diverse applications \citep{barrera23,mccarthy25}. These could also be used to create long pathlength synthetic absorption spectra.

Alternatively, absorption spectra that cover significant cosmological distance i.e. traverse a large redshift interval can be made by combining multiple sightlines from discrete simulation snapshots where pre-computed spectra are available. In particular, we synthesize a long sightline by stitching together many short sightlines across snapshots. We switch between datasets at the midpoints of comoving distance between these snapshots. Each snapshot is responsible for covering a $\Delta z$ centered on its redshift, and an integer $N$ randomly selected spectra are selected to cover the needed pathlength.\footnote{We neglect the continuous $(1+z)^3$ evolution of physical density between snapshots. In addition, an integer number of available spectra do not exactly sum to the required cosmological pathlength, and this can leave wavelength gaps.} The optical depth arrays of each transition are shifted in redshift according to the cumulative distance, interpolated back onto the master (restframe) wavelength grid, accumulated, and ultimately converted back to flux.

\begin{figure*}
   \centering
   \includegraphics[width=0.99\textwidth]{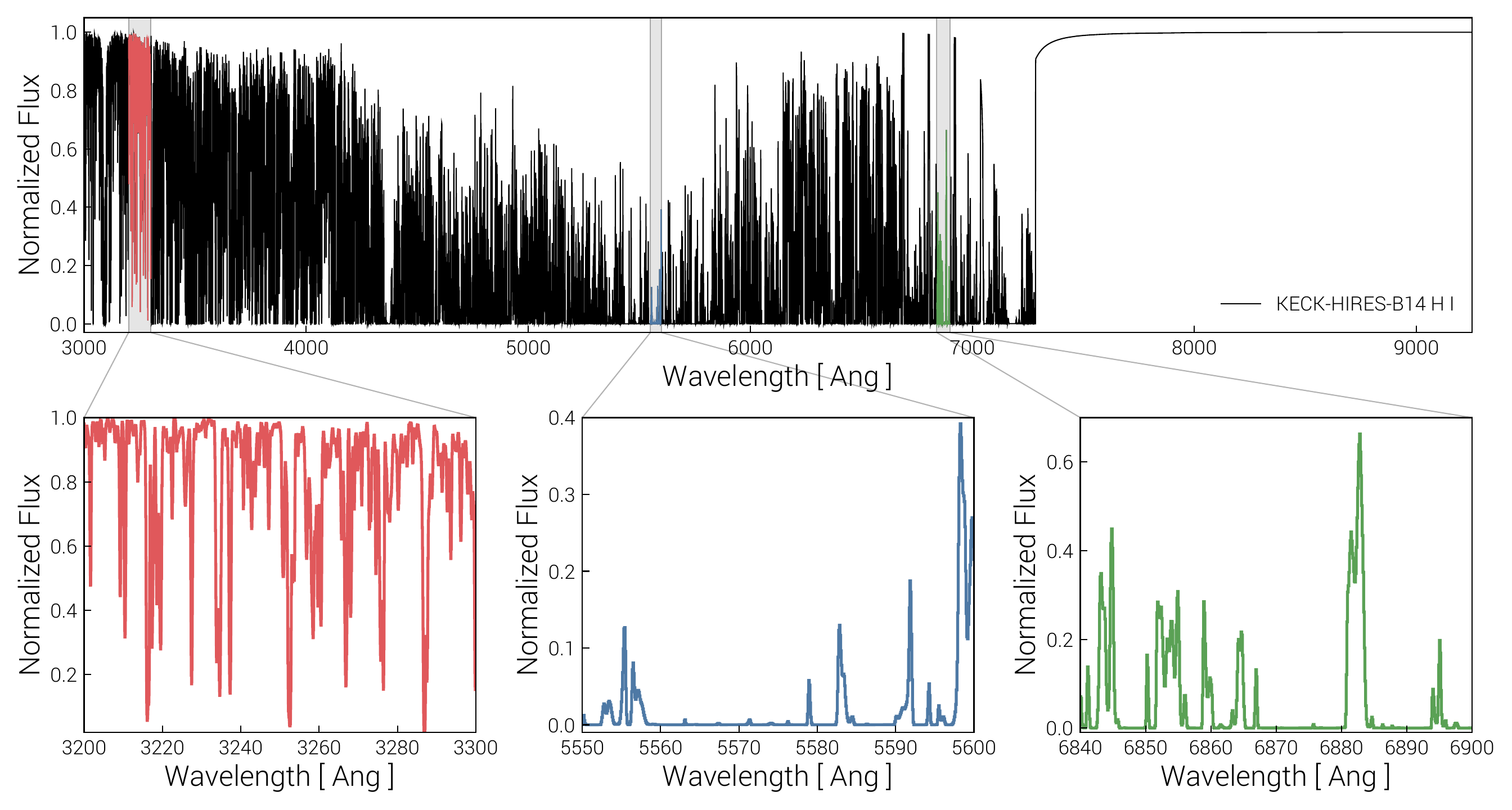}
   \caption{Post-processing of mock absorption spectra to simulate the traversal of a full lightcone i.e. over cosmological distance scales. In this case, we show a synthetic Lyman-$\alpha$ forest spectrum from TNG50, obtained towards a putative background quasar at $z=5.0$. We include all strong transitions of HI, for a KECK-HIRES instrument configuration. The resulting spectrum extends from $3000$\AA\xspace to $9250$\AA\xspace in rest wavelength, shown in the main panel, while below we zoom in on three narrow spectral regions exhibiting predominantly transmission (left) versus absorption (right).}
   \label{fig:lightcone_spec}
\end{figure*}

Figure \ref{fig:lightcone_spec} demonstrates an example of a cosmological-length absorption spectrum made in this way. In particular, we create a KECK-HIRES spectrum towards an imagined background quasar at $z=5$ that includes (only) transitions of hydrogen. In particular, Lyman-$\alpha$ at $\lambda_0 = 1215.670$\AA\xspace dominates, and we also include Ly$\beta$ plus twelve further HI transitions down to $915$\AA. Suppression beyond the Lyman limit is not explicitly included. The main panel shows the full spectrum, while the three insets show zooms into smaller wavelength regions with a variety of structure. At $\sim 3200$\AA\xspace (green) we find predominantly transmission with Ly$\alpha$-forest like dips that frequently reduce the normalized flux to near zero, but rarely to saturation. At $\sim 5550$\AA\xspace (blue) we see mostly absorption, with only occasional transmission spikes \citep{garaldi19}, while $\sim 6900$\AA\xspace (purple) shows an intermediate case with strongly mixed gas phases along the line-of-sight. In the full spectrum, there is a strong damping wing due to neutral hydrogen close to the redshifted transition wavelength $(1 + z) \lambda_0 = 7290$\AA\xspace \citep{miraldaescude98}. Note that we do not include absorption beyond the Lyman limit in this example.

The statistics of the Lyman-alpha forest probe relatively simple physics, predominantly related to the ionizing photon budget of the UVB \citep{gunn65}, and have been used to test cosmological simulations for decades \citep{cen94,hernquist96,theuns98}. They provide a basic sanity check for our spectra. In particular, we compute the mean transmitted flux i.e. Lyman-alpha optical depth using KECK-HIRES like spectra from TNG50. At $z=\{2, 3, 4\}$ we find $\bar{F} = \{0.78, 0.60, 0.32\}$, corresponding to $\tau_{\rm Ly\alpha} = \{0.2, 0.5, 1.1\}$. As a point of comparison, \citet{mcdonald00} find $\bar{F} = \{0.82, 0.68, 0.48\}$ at mean redshifts $\bar{z} = \{2.4, 3.0, 3.9\}$ based on eight high-redshift HIRES quasars, while \citet{ding24} infer $\tau_{\rm Ly\alpha} = \{0.2, 0.6, 1.2\}$ at $z=\{2.5, 3.5, 4.3\}$ using 27k quasars from eBOSS DR14 \citep[see also][]{fg09,paris11,becker13}. The values are roughly consistent, and we anticipate that future work will explore more quantitative comparisons and complex statistics of hydrogen absorption.

Similar long path-length metal-line absorption spectra are useful to make direct comparisons to the methods and results from observational surveys such as XQR-30 \citep{dodorico23}. They are not as necessary in the context of galaxy-centric programs such as MUSE-ALMA Haloes \citep{peroux19} or CUBS \citep{chen20}, where the absorption of interest is localized in redshift. In general, the vast majority of analyses and scientific goals can be realized with relatively short, `single snapshot' sightlines.


\section{Summary and Future Directions} \label{sec_summary}

In this paper we describe the creation and public release of a large library of synthetic absorption spectra from multiple large-scale cosmological hydrodynamical simulations. This resource enables new comparisons between simulations and data, and boosts the utility and interpretability of both. Our deliverables are:

\begin{itemize}
  \item We generate large catalogs of mock (i.e. synthetic) absorption spectra that intersect gas in the ISM of galaxies, their surrounding CGM, and the diffuse IGM.
  \item These spectra are created for multiple underlying cosmological simulations: those of the IllustrisTNG suite, including TNG50 and TNG-Cluster, as well as EAGLE, and SIMBA. They span redshifts from $z=0$ to $z=6$.
  \item Absorption spectra are made for different hydrogen and metal-line transitions, including dozens of ions that span the breadth of observable lines, from HI Ly$\alpha$ to NeVIII 770, OVI 1031, SiIII 1206, CIV 1548, FeII 2631, MgII 2796, NaI 5897, and so on.
  \item Spectra are tailored to the observational characteristics of specific instruments, e.g. SDSS-BOSS, KECK-HIRES, HST-COS, and 4MOST-HRS.
  \item We present a novel algorithm for meshless ray-tracing through Voronoi tessellations and use it to trace sight-lines through simulated gas distributions.
  \item These spectra catalogs are immediately publicly released and made available for use.
\end{itemize}

This dataset can be broadly applied in order to study the relationships between gas physical properties and absorption signatures, quantitatively compare simulations and observations, conduct virtual surveys, and assess observational survey strategies. We also anticipate that the mock spectra, in combination with the underlying simulations, can be used to develop novel tools and methods for the analysis of absorption spectra and the inference of the physical properties of absorbing gas.

A number of past and ongoing publications use our synthetic absorption spectra, including earlier versions and custom datasets tailor-made to replicate specific surveys or instrumental configurations. Table \ref{table_papers} gives an overview, in order to provide illustrative examples and use cases for this dataset.

\subsection{Ongoing and Future Directions}

This is not intended to be a static resource. Instead, \textbf{we welcome suggestions and requests from the community to expand this synthetic absorption spectra library}. These will include adding additional simulations, specialized spectra for new spectrographs and surveys, extra metals or ions, other redshifts, changing modeling assumptions, and so on.

One important future direction is next-generation theoretical models. That is, future cosmological hydrodynamical simulations of galaxy formation. Generally speaking, these will have (i) higher resolution, enabling more small-scale gas structure in diffuse media to be numerically resolved \citep[e.g.][]{ramesh24a,rey24}; (ii) cooling and chemistry below temperatures of $\sim 10^4$\,K, also incorporating models for molecules and dust \citep[e.g.][]{katz22,richings22,maio22}, (iii) new physics such as magnetic fields and cosmic rays, (iv) on-the-fly and/or sophisticated post-processing radiative transfer to improve the ionization and photochemistry modeling \citep[e.g.][]{pallottini19,nakazato25}, and (v) methods for tracking and describing multi-phase structure below the resolution scale \citep{butsky24,smith24,das24}. These improvements, both computational and physical, will undoubtedly manifest in the absorption signatures of our mock spectra.

At the same time, next-generation telescopes and high resolution spectrographs promise to deliver unprecedented observational data with higher sensitivity and spectral resolution. This includes new instrumentation on the Extremely Large Telescope \citep[ELT/ANDES;][for example]{marconi24}. Massively multiplexed spectrographs will push beyond SDSS/DESI type statistical inferences \citep[e.g. WEAVE and 4MOST;][in addition to more ambitious and distant successors]{piere16,peroux23}. New windows into the CGM/IGM at previously inaccessible wavelengths are likewise enabled by new instruments \citep[e.g. XRISM, HUBS, and NewAthena in the X-ray;][]{xrism20,bregman23,cruise25}. 

As simulations and observations advance, statistically robust libraries of synthetic absorption sightlines will become increasingly powerful tools to shed light on the physics underlying the formation and evolution of galaxies, and the associated flows of gas and metals, across cosmic time.

\subsection{Caveats and Important Considerations}

Several important assumptions and limitations exist and should be carefully considered depending on the scientific use case at hand. These are either inherent to the hydrodynamical simulations that provide our underlying ground truth for the properties and distribution of gas, or in the post-processing analysis needed to derive mock absorption spectra.

{\renewcommand{\arraystretch}{1.4}
\begin{table*}
    \centering
    \begin{tabular}{p{0.24\linewidth} p{0.29\linewidth} p{0.22\linewidth} p{0.15\linewidth}}
        \hline\hline
        Paper & Topic & Instrument(s), Survey(s) & Simulation(s)\\
        \hline\hline
        \textcolor{blue}{Nelson (2026)} & this paper & multiple & multiple \\ \hline
        \textcolor{blue}{Guerra Varas et al. (in prep)} & MgII with ML & 4MOST, ByCycle & TNG50 \\
        \textcolor{blue}{Wu et al. (in prep)} & MgII cosmic incidence in ASPIRE & JWST/NIRCam WFSS & TNG50 \\
        \textcolor{blue}{Guo et al. (in prep)} & Milky Way sightline pairs & HST/COS & GIBLE \\
        \textcolor{blue}{Boucsein et al. (in prep)} & Ly$\alpha$-forest cosmological inference & SDSS & CLIMB \\
        \textcolor{blue}{\citet{kong26}} & MgII vs CIV in ELGs and QSOs & DESI & TNG50 \\
        \textcolor{blue}{\citet{apariciomarcos26}} & coronal broad Lyman-$\alpha$ & HST/COS & TNG50 \\
        \textcolor{blue}{\citet{singh26}} & OVI in groups & HST/COS, COS-IGrM & TNG50, SIMBA \\
        \textcolor{blue}{\citet{santos25}} & strongly blended Lyman-$\alpha$ & SDSS/BOSS, KECK/HIRES & TNG50 \\
        \textcolor{blue}{\citet{guo25}} & FRB dispersion measure + ML & ray-tracing & CAMELS-TNG \\ 
        \citet{szakacs23} & identifying MgII with ML & 4MOST, ByCycle & TNG50 \\
        \hline
    \end{tabular}
    \caption{$\vphantom{X^{X^{X}}}$ List of published and ongoing work that uses the synthetic absorption spectra generator and/or catalogs.}
    \label{table_papers}
\end{table*}
}

\subsubsection{Radiation and Photochemistry}

One major challenge is the accurate modeling of the radiation fields that permeate the CGM and IGM. This requires realistic results for the distribution of sources (stars and AGN), as well as methods to solve for the propagation of this radiation through of the intervening gas. None of the simulations we currently analyze include on-the-fly radiative transfer (RT). As a result, spatially-dependent radiative processes are either neglected or only approximately accounted for.\footnote{In the case of the TNG model, for example, the uniform UVB is attenuated with an on-the-fly self-shielding model, while the radiation field from AGN affects gas with a simple $1/r^2$ optically thin treatment.}

We do not post-process simulations with any form of RT. This means that we adopt the current state of gas, including its temperature, as inputs to our CLOUDY post-processing. While a reasonable assumption far from sources of radiation, this undoubtedly becomes inaccurate near galaxies \citep{zhu24}. In particular, we do not include `local' radiation sources of any kind in the CLOUDY simulations. Caution should be used for sightlines that pass near, or even within, the star-forming bodies of galaxies, near bright AGN, and so on. Overall, we expect gas to be more ionized in these environments \citep{suresh17,holguin25}.

By adopting the current instantaneous properties of gas, we also by definition neglect any time-dependence i.e. non-equilibrium effects. Gas that has recently undergone a shock, has high levels of turbulence \citep{buie20}, or that has experienced a highly time-variable incident radiation field \citep{oppenheimer18a}, may be out of equilibrium. In addition, at low densities in the outer CGM or IGM, long cooling times can easily lead to out of equilibrium states \citep{cen06}. Our CLOUDY calculations assume collisional plus photoionizaton equilibrium, and so do not capture these effects. Incorporating such processes would shift the abundances of different ions in a complex, time and environment-dependent way.

\subsubsection{The UVB}

Relatedly, most cosmological simulations not focusing on the epoch of reionization assume that the UVB is a spatially uniform, redshift and frequency dependent radiation field. In creating our mock spectra, our treatment of the UVB incorporates a simple self-shielding model, as a function of gas density \citep{rahmati13}. This is a significant improvement over assuming that the UVB permeates space in the optically thin limit, but is still a simplified approach. For example, it does not integrate the contribution to the shielding column outside of a particular parcel of gas, and so cannot capture effects that accumulate due to complex gas geometries \citep{baumschlager24}.

The inclusion of the UVB also makes our results dependent on the particular UVB chosen \citep{fg09}, and different UVBs can lead to non-negligible differences for metal ionization states \citep{taira25}.

\subsubsection{Resolution}

Ultimately, the realism of the mock absorption spectra is linked to the realism of the underlying hydrodynamical simulation(s). Users should be aware of the physical and computational challenges of modeling multi-phase and multi-scale gas. In particular, density structure -- such as cool clouds -- and kinematic structure -- such as turbulence -- are only resolved down to the resolution limit of a given simulation. Absorption tracing cool-dense gas, such as HI and OI, will in general be more difficult for cosmological simulations to capture; in contrast, tracers of volume-filling warm/hot phases may be more robust.

In addition, models that use effective multi-phase ISM models \citep[e.g.][true of most current large-volume cosmological simulations]{springel03} are particularly limited within star-forming gas. On one hand, spectra from these simulations inherit these limitations, and next-generation models will improve the utility of synthetic spectra as interpretative tools. On the other hand, galaxy formation simulations will always invoke sub-grid models below the resolution scale, and the sensitivity of absorption signatures to physical model details allows us to assess, constrain, and improve these models.

\subsubsection{Metal Abundances and Stellar Yields}

Most modern cosmological hydrodynamical simulations include models for stellar evolution and the resulting production of heavy elements. Some subset of metal abundances are then tracked, in stars and in the gas-phase they return to. In the TNG model, for example, C, N, O, Ne, Mg, Si, and Fe are followed individually, in addition to total metallicity. This information reflects the complex interplay of metal production from different channels, i.e. from supernova Ia, II, and stellar winds, as well as subsequent advection and mixing in the gas-phase. This treatment enables science applications beyond the assumption of solar abundance patterns \citep[e.g.][]{kumar24}.

However, tracking metal abundances makes results sensitive to the particular stellar yield models adopted, and significant uncertainties exist therein \citep{pillepich18a}. It also introduces additional differences when post-processing simulations with different yield tables, and/or numerical hydrodynamical techniques that can strongly influence metal mixing \citep{shah25}. Our fiducial choice is to use metal abundances when available, and to assume solar abundance ratios otherwise.

\subsection{Future Additions to the Mock Spectra Catalogs}

A number of specific improvements and expansions are planned for the future:

\begin{enumerate}
\item We discuss the use of dust depletion models to correct (i.e. reduce) gas phase abundances of some metals. These will be available in the future, keeping in mind that their results will depend on assumptions and choices related to dust composition and abundance.

\item The ionization state of gas is sensitive to the choice of UVB, and we currently provide spectra for only our one fiducial UVB choice. In the future we can vary the UVB and assess its impact.

\item The use of directly tracked simulation metal abundances introduces simulation-dependent uncertainty. Spectra instead made by (always) assuming solar abundance ratios avoid this issue.

\item Despite the large numbers of sightlines, statistics are still limited, particularly for strong i.e. rare absorbers. We currently focus on a setup of one million equally-spaced sightlines covering each simulation snapshot of interest, along a single ($\hat{z}$) line-of-sight direction. These will be expanded to a larger configuration, of four million randomly located sightlines along the three orthogonal directions of each snapshot.

\item Some experiments may benefit from ultra-dense sightline sampling, or require more statistics at low impact parameters around halos or galaxies with specific properties, and so on. These can be created when useful.

\item Realistic, instrument-specific noise models and templates will be critical for detailed analyses. At present, our examples in this presentation typically include only Gaussian random noise, and it is left to the user to apply more realistic noise, if needed for a particular science application, on top of the provided spectra.
\end{enumerate}


\section*{Data Availability}

The main result of this paper is a data product available at \url{www.tng-project.org/spectra}. This is a (very) large library of $\sim\,$1 billion publicly downloadable synthetic absorption spectra, covering a wide variety of ions, transitions, instruments, observational characteristics, redshift ranges, and derived from multiple numerical simulations. This catalog of mock spectra is dynamic and will grow over time.

All the IllustrisTNG simulations including TNG50 are publicly available and accessible at \url{www.tng-project.org/data}, as described in \cite{nelson19a}.

The Voronoi mesh ray-tracing and mock absorption spectra generation software are also public as part of the \textsc{temet} library,\footnote{\url{www.github.com/dnelson/temet}} as linked from the SALSA data page.


\section*{Acknowledgements}

DN thanks Chris Byrohl, Reza Ayromlou, and the other members of the group in Heidelberg throughout 2023 and 2024 who gave comments and suggestions on many early directions; Roland Szakacs, Qi Guo, Duarte Mu\~{n}oz Santos, Patricia Marcos, Tanmay Singh, Nicolas Guerra Varas, Simon Weng, Kirill Tchernyshyov, and Jeremias Boucsein for exploring early versions of these datasets; and Annalisa Pillepich for comments and ideas throughout the preparation of this paper.

This research was supported by the International Space Science Institute (ISSI, \url{https://www.issibern.ch/}) in Bern, through ISSI International Team project \#564 (The Cosmic Baryon Cycle from Space).
DN acknowledges funding from the Deutsche Forschungsgemeinschaft (DFG) through an Emmy Noether Research Group (grant number NE 2441/1-1). This work is supported by the Deutsche Forschungsgemeinschaft (DFG, German Research Foundation) under Germany's Excellence Strategy EXC 2181/1 - 390900948 (the Heidelberg STRUCTURES Excellence Cluster). S.L. acknowledges support by FONDECYT grant 1231187. Our calculations have primarily made use of the Vera cluster of the Max Planck Institute for Astronomy (MPIA), operated by the Max Planck Computational Data Facility (MPCDF).

\bibliographystyle{aa}
\bibliography{refs}

\appendix

\section{Meshless Voronoi Ray-Tracing} \label{sec:voronoi}

\begin{algorithm}
\DontPrintSemicolon
\caption{Meshless Voronoi ray-tracing (see text).}\label{alg:trace}
\KwIn{$\mathcal{T}$, $\vec{x_i}$, $\vec{r_0}$, $\hat{r}$, $S$}
\KwOut{$\delta x$, $\mathcal{I}$}

\BlankLine
$dl = 0.0$, 
$\vec{r} = \vec{r_0}$, 
$\vec{r}_{\rm f} = \vec{r_0} + \hat{r} \cdot S$\;
$I_{\rm cur}$ = \textsc{tree\_nearest}($\vec{r_0}, \mathcal{T})$, 
$I_{\rm f}$ = \textsc{tree\_nearest}($\vec{r_{\rm f}}, \mathcal{T})$\; 
$\mathcal{S}_{\rm I, dist}$ = [] \comment{\# stack of index, distance pairs}\;
\BlankLine

\While(\comment{\# still intersecting cells}){not finished}{
  $\vec{x}_{\rm cur} = \vec{x_i}[I_{\rm cur}]$\; 
  $I_{\rm end} = -1$, 
  $L = 0.0$, 
  $R = S - dl$, 
  $dl_{\rm local} = \infty$\; 
  
  \uIf(\comment{\# use closest failed dist as start}){$\mathcal{S}$ not empty}
  {
    $\textsc{stack\_remove\_all}(\mathcal{S}_{\rm I}, I_{\rm cur})$ \comment{\# avoid self}\;
    $R = 2 \times \textsc{stack\_pop}(\mathcal{S}_{\rm dist})$\;
    $I_{\rm end} = \textsc{stack\_pop}(\mathcal{S}_{\rm I})$\;
  }

  \BlankLine
  \For(\comment{\# while not ending inside a neighbor}){$n$++}{
    $l_{\rm cen} = (L + R) / 2$ \comment{\# new test position along ray}\; 
    $\vec{r}_{\rm end} = \vec{r} + \hat{r} \cdot l_{\rm cen}$\; 

    \uIf{$n > 0$ or $I_{\rm end} == -1$}
    {
      $I_{\rm end} = \textsc{tree\_nearest}(\vec{r}_{\rm end}, \mathcal{T})$\;
    }

    \If(\comment{\# same as current cell}){$I_{\rm end} == I_{\rm cur}$}
    {
      \eIf(\comment{\# skipped neighbor, bisect}){$R - L > \epsilon$}
      { $L = (L + R) / 2$\; }
      (\comment{\# close to face, but yet to exit}){ $R \pluseq \epsilon$\; }
      continue\;
    }

    $\vec{x}_{\rm end}$ = $\vec{x_i}[I_{\rm end}]$ \comment{\# position of parent cell}\;

    $dl_{\rm local} = \textsc{intersect}(\vec{x}_{\rm cur}, \vec{x}_{\rm end}, dl_{\rm local})$\;

    $\vec{r}_{\rm cand} = \vec{r} + \hat{r} \cdot dl_{\rm local}$ \comment{\# candidate ray position}\; 

    $I_{\rm cand}$ = \textsc{tree\_nearest}($\vec{r}_{\rm cand}, \mathcal{T}$)\; 

    \If(\comment{\# false intersection}){$I_{\rm cand} \not\in [I_{\rm cur}, I_{\rm end}]$}
    {
      $l_{\rm search} = \left( \vec{r}_{\rm cand} - \vec{r} \right) \cdot \hat{r}$ \comment{\# next candidate}\;
      \uIf{$l_{\rm search} > L$}
      { $R = 2 \times l_{\rm search} - L$ \comment{\# set to new (L+R)/2}\; } 

      \textsc{stack\_append}($I_{\rm cand}, l_{\rm cen} + \epsilon$) \comment{\# add failure}\;

      continue \comment{\# neighbor incorrect, search again}\;
    }

    \comment{\# found a natural neighbor, the correct next cell}\;
    $dl \pluseq dl_{\rm local}$\;
    $\vec{r} \pluseq \hat{r} \cdot dl_{\rm local}$\;

    $\mathcal{S}_{\rm dist} \minuseq dl_{\rm local}$ \comment{\# update max dists for next}\;
    
    \textsc{save\_intersection}$(I_{\rm cur}, dl_{\rm local})$\;

    $I_{\rm cur} = I_{\rm end}$\;

    \If(\comment{\# next cell is the last}){$I_{\rm cur} == I_{\rm f}$}
    {
      $dl_{\rm local} = S - dl$ \comment{\# remaining ray path-length}\;
      $dl \pluseq dl_{\rm local}$\;
      $\vec{r} \pluseq \hat{r} \cdot dl_{\rm local}$\;
      \textsc{save\_intersection}$(I_{\rm cur}, dl_{\rm local})$\;
      finished = True \comment{\# entire ray is done}\;
    }

    break \comment{\# terminate bisection search, move to next}\;
  }

}
\end{algorithm}

\begin{algorithm}
\DontPrintSemicolon
\caption{Helper function \textsc{intersect()} to identify closest valid intersection between a ray and a face.}\label{alg:intersect}
\KwIn{$\vec{r}$, $\hat{r}$, $\vec{x}_{\rm cur}$, $\vec{x}_{\rm end}$, $dl_{\rm local}$}
\KwOut{$dl_{\rm local}$ \comment{\# change indicates valid intersection}}

$\vec{m} = (\vec{x}_{\rm cur} + \vec{x}_{\rm end}) / 2$ \comment{\# edge midpoint on Voronoi face, but only if end cell is an actual neighbor}\;

$\vec{c} = \vec{m} - \vec{r}$ \comment{\# vector from the current ray position to m}\;

$\vec{q} = \vec{x}_{\rm end} - \vec{x}_{\rm cur}$ \comment{\# normal to face plane}\;

\eIf(\comment{\# test intersection of ray and face plane}){$\vec{c} \cdot \vec{q} > 0$}
{ $s = \vec{c} \cdot \vec{q} / \hat{r} \cdot \vec{q}$ \comment{\# standard case, length to intersection}\; }
{ \eIf(\comment{\# point on wrong side of face}){$\hat{r} \cdot \vec{q} > 0$}
  { $s = 0$ \comment{\# direction is away from cell}\; }
  { $s = \infty$ \comment{\# direction is into cell (ignore)}\; }
}

\If(\comment{\# valid, closest, intersection}){$s \geq 0$ and $s \leq dl_{\rm local}$}
{ $dl_{\rm local} = s$ \comment{\# best candidate for next neighbor}\; }

return $dl_{\rm local}$\;
    
\end{algorithm}

Algorithm \ref{alg:trace} gives the pseudo-code of the approach \citep[see also][who recently discuss a similar algorithm for integrations of FRB dispersion measure]{konietzka25}. Its inputs are a pre-constructed tree $\mathcal{T}$, the array of Voronoi cell positions $\vec{x_i}$, the ray starting position $\vec{r_0}$, the ray direction unit vector $\hat{r}$, and the total ray pathlength $S$. The \textsc{save\_intersection} function builds the two outputs: ordered lists of intersected cell path-lengths $\delta x$ and cell indices $\mathcal{I}$. The \textsc{tree\_nearest} function returns the index of the closest cell using $\mathcal{T}$ (details omitted). We also omit details related to handling of periodic boundary conditions. Note that the stack $\mathcal{S}$ of cell index and distance pairs holds previous failed candidates, and only accelerates the bisection searches, but is not strictly required. Finally, \textsc{intersect} is a helper function that performs the geometrical ray-Voronoi face plane intersection test, as shown in Algorithm \ref{alg:intersect}. 

In practice, the tree is first constructed and then the above function is called in a loop over rays within a shared memory, thread parallel context. The performance of this algorithm is excellent. We also develop, and compare to, a traditional ray-tracing algorithm, that has the following steps: (i) explicitly construct the global Voronoi mesh, (ii) obtain the natural neighbor connectivity information of each cell, (iii) perform ray-plane intersection against each neighbor face, (iv) identify the closest positive intersection to determine the pathlength in the current cell as well as the next cell, (v) iterate.

Even neglecting the mesh construction cost of the first step, which is more expensive than constructing the octtree needed for the meshless algorithm, we find that the meshless algorithm is actually much faster. In practice, the speed-up can be a factor of several: roughly 6x in the case of full-box rays through TNG50-4. We also compare the outcomes of the meshless and full mesh algorithms to verify the correctness of our new approach.

\section{Observational Realism}

\begin{figure*}
   \centering
   \includegraphics[width=0.46\textwidth]{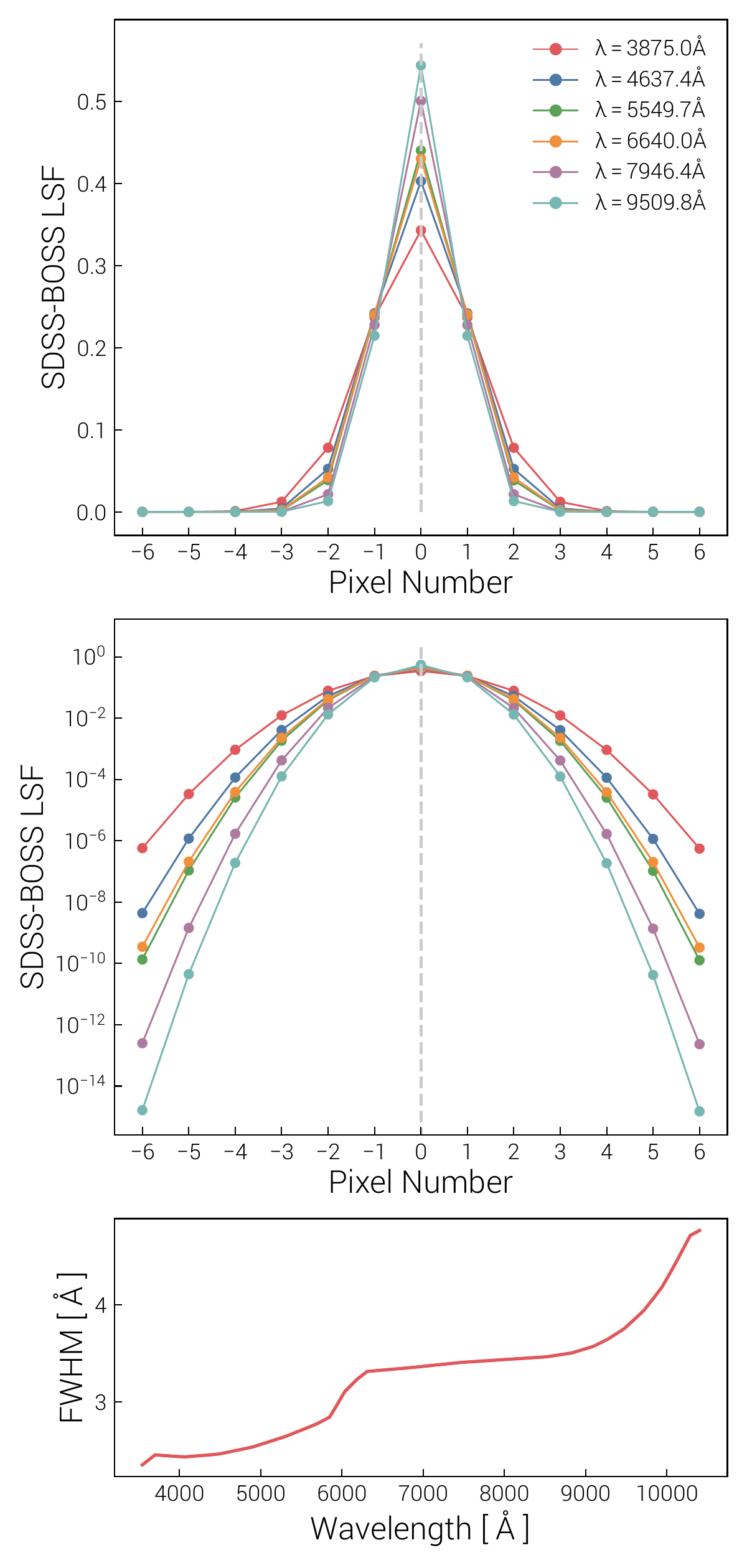}
   \includegraphics[width=0.46\textwidth]{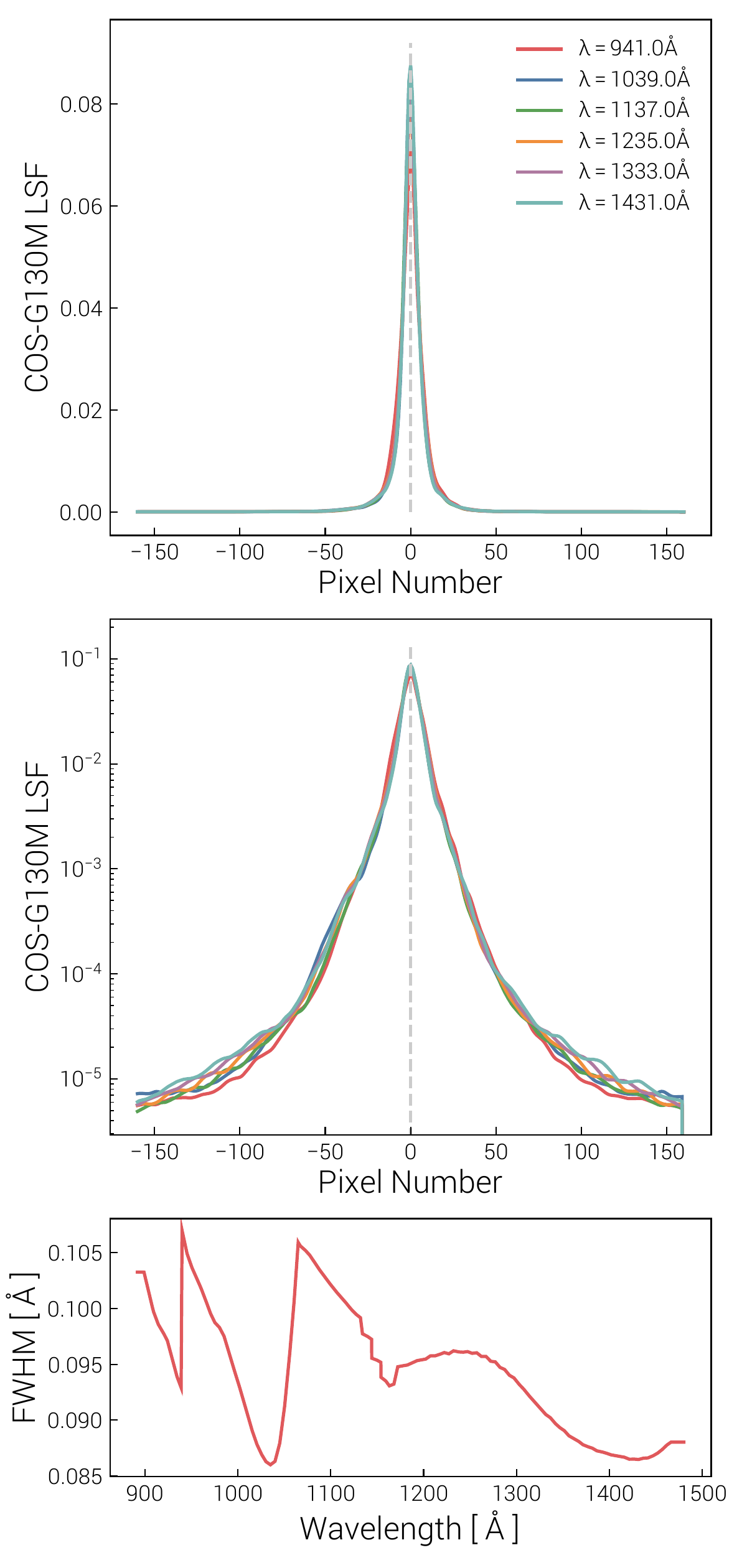}
   \caption{Line spread function (LSF) characterization of the finite spectral resolution available with observational spectrographs. Left: an example showing the multi-object fiber-fed spectrograph of the Baryon Oscillation Spectroscopic Survey of the Sloan Digital Sky Survey \protect\citep[SDSS-BOSS;][]{smee13}. Results for spectrograph 1, where spectrograph 2 is worse at $R \sim 100$ at the blue end. The LSF in this case is measured from calibration arc images taken before each set of science exposures, and is based on a sample of 100 BOSS plates. Right: showing the LSF of the G130M grating on HST/COS.}
   \label{fig:lsf}
\end{figure*}

Figure \ref{fig:lsf} shows the line spread function (LSF; top two panels) and spectral resolution as characterized by the FWHM (bottom panel) for the SDSS-BOSS instrument (left). As described in our methodology, we generally create instrument-specific spectra that include wavelength-dependent convolution by the observation LSF. This can be an important ingredient in the comparison of models and data.

The right panels of Figure \ref{fig:lsf} show a similar example for the G130M grating of HST/COS. Here the well known and empirically characterized LSF has significant and extended wings.

Note that these are two examples of adding finite spectral resolution characteristics to our mock spectra, as one component of observational realism. We also provide the `idealized' datasets that do not apply any LSF, so that users can customize this part of the forward modeling as needed.

\end{document}